

\tolerance=3000 \hbadness=3000 \hfuzz=6pt
\magnification=1200
\baselineskip=15pt plus 2pt minus 1pt
\font\sevteenrm=cmr17
\font\twelverm=cmr12
\font\eightrm=cmr8
\def\rm{\fam0 \tenrm}
\overfullrule=0pt



\newcount\yearltd\yearltd=\year\advance\yearltd by -1900
{\count255=\time\divide\count255 by 60 \xdef\hourmin{\number\count255}
  \multiply\count255 by-60\advance\count255 by\time
  \xdef\hourmin{\hourmin:\ifnum\count255<10 0\fi\the\count255}}

\newwrite\ffile  
\newif\iffclosed \fclosedtrue 
\def\fwrite#1{\iffclosed\immediate\openout\ffile=\jobname.fwd\fclosedfalse\fi
              \immediate\write\ffile{#1}}


\def\freference#1{\fwrite{\noexpand\xdef\noexpand#1\noexpand{#1}} }

\newread\tfile
\def\testinput#1{
                 \immediate\openin\tfile=#1
                 \immediate\ifeof\tfile\else\closein\tfile\input#1\fi
                 \closein\tfile
                  }



%


\def\eqlabform#1{{\noexpand\ifmmode\hbox{$(\chapsyme\secsyme#1)$}
	          \noexpand\else Eq.\ $\chapsyme\secsyme#1$\noexpand\fi}}

\def\thmlabform#1{{$\chapsyme\secsyme#1$}}
\def\figlabform#1{{$#1$}}

\def\nolabels{\def\eqnlabel##1{}\def\eqlabel##1{}\def\reflabel##1{}}
\def\writelabels{\def\eqnlabel##1{%
{\escapechar=` \hfill\rlap{\hskip.09in\string##1}}}%
\def\eqlabel##1{{\escapechar=` \rlap{\hskip.09in\string##1}}}%
\def\reflabel##1{\noexpand\llap{\string\string\string##1\hskip.31in}}}
\nolabels

\def\eqname#1{\global\advance\meqno by1
               \xdef #1{\eqlabform{\the\meqno}} }
\def\eqaname#1{\global\advance\meqno by1
               \xdef #1##1{{\eqlabform{\the\meqno##1}}} }

\def\eqna#1{\eqaname#1\eqnlabel{#1$\{\}$}}
\def\eqnn#1{{\eqname#1\eqnlabel#1}}

\def\etag#1{\eqname#1\eqno#1\eqnlabel#1}

\def\eqnt#1{\global\advance\meqno by1 \xdef #1##1{\hbox{$(T\the\meqno##1)$}}%
            \eqnlabel{#1$\{\}$}}

\def\ftag#1{\global\advance\meqno by1
            \xdef #1{Eq. \chapsyme\secsyme\the\meqno}
            \eqno\hbox{$(\chapsyme\secsyme\the\meqno)$}
            \freference{#1} }

\def\newthm#1{\global\advance\thmno by
1\xdef#1{{\thmlabform{\the\thmno}}}#1\freference{#1}}
\def\newfig#1#2#3{
\global\advance\figno by 1\xdef#1{{\figlabform{\the\figno}}}\freference{#1}
\bigskip
\psfig{file=#2}
\nobreak\medskip
\noindent{\bf Figure #1.  }{\it  #3}
\medskip
}

\catcode `@=11
\newtoks\@temptokena  

\newif\if@afterindent \@afterindenttrue
\newdimen\@tempdima
\def\contentdepth{1}
\def\@pnumwidth{20pt}
\def\@tocrmarg{80pt}
\def\@dotsep{1.7}

%
%
\def\dottedtocline#1#2#3#4#5{\ifnum #1>\contentdepth \else
  \vskip \z@ plus .2pt
  {\leftskip #2\relax \rightskip \@tocrmarg \parfillskip -\rightskip
    \parindent #2\relax\@afterindenttrue
   \interlinepenalty\@M
   \leavevmode
   \@tempdima #3\relax \advance\leftskip \@tempdima \hbox{}\hskip -\leftskip
   #4\nobreak\leaders\hbox{$\m@th \mkern \@dotsep mu.\mkern \@dotsep
       mu$}\hfill \nobreak \hbox to\@pnumwidth{\hfil\rm #5}\par}\fi}
\def\numberline#1{\hbox to\@tempdima{#1\hfil}}

\xdef\chapsym{}\xdef\chapsyme{} 
\xdef\secsym{}\xdef\secsyme{}
\global\newcount\chapno \global\chapno=0
\global\newcount\secno \global\secno=0
\global\newcount\subsecno \global\subsecno=0
\global\newcount\meqno \global\meqno=0
\global\newcount\thmno \global\thmno=0
\global\newcount\figno \global\figno=0
\global\newcount\ftno \global\ftno=0
\def\ourfolio{\ifnum\pageno<0 \romannumeral-\pageno \else\number\pageno\fi}
\def\leaderfill{\leaders\hbox to 1em{\hss . \hss}\hfill}

\newwrite\cfile  
\newif\ifcclosed \cclosedtrue 
\def\cwrite#1{\ifcclosed\immediate\openout\cfile=\jobname.con\cclosedfalse\fi
              \immediate\write\cfile{#1}}

\newwrite\pfile  
\newif\ifpclosed \pclosedtrue
\newcount\pno   \pno=0  
\def\pwrite#1{\ifpclosed\immediate\openout\pfile=\jobname.pre\pclosedfalse\fi
              \immediate\write\pfile{#1}}

\def\extrahead#1{\vfil\eject\centerline{#1}\bigskip}

\def\extra#1{\extrahead{#1} \@temptokena{#1}
  \cwrite{\noexpand \dottedtocline{0}{0pt}{20pt}
  {\the\@temptokena}{\ourfolio} } }
\def\extraearly#1{\extrahead{#1}\@temptokena{#1}
  \pwrite{\noexpand \dottedtocline{0}{0pt}{20pt}
  {\the\@temptokena}{\ourfolio}} }

\def\chapter#1#2#3{
\global\ftno=1
\xdef#1{#2}
\xdef\chapsym{#2}
\xdef\chapsyme{#2.}
\global\thmno=0\global\figno=0\global\meqno=0\global\secno=0\global\subsecno=0
\vfill\eject
\noindent{\bf \chapsyme\ #3}
\@temptokena{#3}
\cwrite{\noexpand
        \dottedtocline{0}{0pt}{15pt}
           {\noexpand\numberline{#2.}{\the\@temptokena}}{\ourfolio}}
\freference{#1}
\par\nobreak\medskip\nobreak
}

\def\newchap#1#2{\global\advance\chapno by 1
                 \chapter#1{\the\chapno}{{#2}}
                 }

\def\section#1#2#3{
\xdef\secsym{\chapsyme#2}
\xdef\secsyme{\secsym.}
\xdef#1{{\secsym}}
\global\subsecno=0
\bigbreak\medskip
\@temptokena{#3}
\noindent{\bf\secsyme\ #3 }
\cwrite{\noexpand
        \dottedtocline{1}{15pt}{25pt}
              {\noexpand\numberline{\secsyme}{\the\@temptokena}}{\ourfolio}}
\freference{#1} }
\def\newsec#1#2{\global\advance\secno by 1
                 \section#1{\the\secno}{{#2}}
                 }

\def\subsection#1#2#3{
\xdef#1{{\secsyme#2}}
\medbreak\smallskip
\noindent{\it\secsyme#2 #3 }
\@temptokena{#3}
\cwrite{\noexpand
        \dottedtocline{2}{40pt}{25pt}
           {\noexpand\numberline{\secsyme#2.}{\the\@temptokena}}{\ourfolio}}
\freference{#1}
}
\def\newsubsec#1#2{\global\advance\subsecno by 1
                   \subsection#1{\the\subsecno}{{#2}}}

\def\subsubsec#1{{\medbreak\smallskip\noindent{\it #1}}}

\def\listtoc{\vfill\eject
             \immediate\closeout\pfile
             \centerline{{\bf Table of Contents}}\bigskip
             \testinput{\jobname.pre}
             \testinput{\jobname.con}
             \vfill\eject
              }

\def\footfont{\eightrm}
\newskip\footskip\footskip12pt plus 1pt minus 1pt 
\def\f@@t{\footfont\baselineskip\footskip\bgroup\aftergroup\@foot\let\next}
\setbox\strutbox=\hbox{\vrule height9.5pt depth4.5pt width0pt}
\def\foot{\global\advance\ftno by1\footnote{$^{\the\ftno}$}}
%
\newwrite\ftfile
\def\footend{\def\foot{\global\advance\ftno by1\chardef\wfile=\ftfile
$^{\the\ftno}$\ifnum\ftno=1\immediate\openout\ftfile=foots.tmp\fi%
\immediate\write\ftfile{\noexpand\smallskip%
\noexpand\item{f\the\ftno:\ }\pctsign}\findarg}%
\def\footatend{\vfill\eject\immediate\closeout\ftfile{\parindent=20pt
\centerline{\bf Footnotes}\nobreak\bigskip\input foots.tmp }}}
\def\footatend{}

\catcode `@=12


\global\newcount\refno \global\refno=1
\newwrite\rfile

\def\ref#1#2{
    \nref#1{#2}}
\def\nref#1#2{\xdef#1{{\the\refno}}%
\ifnum\refno=1\immediate\openout\rfile=\jobname.ref\fi%
\immediate\write\rfile{\noexpand\item{[#1]\ }\reflabel{#1}#2}%
\global\advance\refno by1}
\def\addref#1{\immediate\write\rfile{\noexpand\item{}#1}}

\def\semi{;\hfil\noexpand\break}

\def\listrefs{\immediate\closeout\rfile
              \parindent=20pt\baselineskip=14pt
              \extra{References}
              {\frenchspacing%
               \testinput{\jobname.ref}\vfill\eject}
              \nonfrenchspacing
              }


\def\book#1#2#3{{\it #1}, #2, #3}


\def\article#1#2#3#4#5{{\it #1}, #2 {\bf #3} (#4) #5}

\def\preprint#1#2{{\it #1}, #2}



\def\Cal{\cal}
\def\bold#1{{{{\bf #1}}}}
\def\cal{\fam2 } \def\rm{\fam0 \tenrm} 

\def\Bf{\bold}



\def\cA{{\Cal{A}}} \def\cB{{\Cal{B}}} \def\cC{{\Cal{C}}} \def\cD{{\Cal{D}}}
   
\def\cI{{\Cal{I}}} \def\cJ{{\Cal{J}}}  \def\cL{{\Cal{L}}}
 \def\cN{{\Cal{N}}} \def\cO{{\Cal{O}}} 
 \def\cR{{\Cal{R}}}

\def\IR{{\hbox{{\rm I}\kern-.2em\hbox{\rm R}}}}
\def\IB{{\hbox{{\rm I}\kern-.2em\hbox{\rm B}}}}
\def\IC{{\ \hbox{{\rm I}\kern-.6em\hbox{& ams\bf C}}}}
\def\IQ{{\ \hbox{{\rm I}\kern-.54em\hbox{\bf Q}}}}
\def\IZ{{\hbox{{\rm Z}\kern-.4em\hbox{\rm Z}}}}

\def\Sum{\sum}   
\def\union{\bigcup}
\def\half{{ 1 \over 2}}

\def\del{\partial}

\def\part#1{{\partial\over\partial #1}}

\def\dirac{{\hbox{$\partial$\kern-.53em$/$}}}

\catcode`\@=11
\def\FN@{\futurelet\next}
\def\DN@{\def\next@}
\def\relaxnext@{\let\next\relax}
\def\nolimits@{\relaxnext@
 \DN@{\ifx\next\limits\DN@\limits{\nolimits}\else
  \let\next@\nolimits\fi\next@}%
 \FN@\next@}
\def\newmcodes@{\mathcode`\'"27\mathcode`\*"2A\mathcode`\."613A%
 \mathcode`\-"2D\mathcode`\/"2F\mathcode`\:"603A }
\def\operatorname#1{\mathop{\newmcodes@\kern\z@\fam\z@#1}\nolimits@}
\catcode`\@=\active
\def\op{\operatorname}

\def\Im{{  \op{ Im } }}  
  
  \def\Tr{{  \op{Tr  } }}
 \def\ad{{  \op{ad  } }} 
 \def\Ker{{ \op{ Ker} }} 
 


\ref\ALR{L.\ Alvarez-Gaume, J.M.F.\ Labastida, and A.V.\ Ramallo,
 \article{A Note on Perturbative Chern--Simons Theory}
   {Nucl. Phys.}{B334}{1990}{103}.}

%

\ref\APS{M.\ Atiyah, V.\ Patodi and I.M.\ Singer,
 {\it Spectral asymmetry and Riemannian geometry } I
   {Math. Proc. Camb. Phil. Soc. } {\bf 77}, 43--69 (1975),
 and II {\bf 78 } 405--432 (1975).}

\ref\AS{S.\ Axelrod and I.\ M.\ Singer,
 \preprint{Perturbative Chern--Simons Quantum Field Theory}{in preparation}.}

\ref\BN{D.\ Bar-Natan,
 \preprint{Perturbative Chern-Simons theory}{Princeton preprint};
 \book{Perturbative Aspects of the Chern-Simons Topological Quantum Field
   Theory}{Princeton University Ph.D. Thesis}{1991}.}

\ref\BNW{D.\ Bar-Natan and E.\ Witten,
 \preprint{Perturbative Expansion of Chern--Simons Theory with
   Non-compact Gauge Group} {IAS preprint HEP-91/4}.}

\ref\BR{D.\ Birmingham and M.\ Rakowski,
 \article{Superfield Formulation of Chern--Simons Supersymmetry}
  {Mod. Phys. Lett.}{A4}{1989}{1753}.}

\ref\BRT{D.\ Birmingham, M.\ Rakowski, and G.\ Thompson,
 \article{Renormalization of Topological Field Theory}
  {Nucl. Phys.}{B329}{90}{83}.}

\ref\BC{A.\ Blasi and R.\ Collina,
 \article{Finiteness of the Chern-Simons Model in Perturbation Theory}
   {Nucl. Phys.}{B345}{1990}{472-492}.}

\ref\BT{R.\ Bott and L.W.\ Tu,
  \book{Differential Forms in Algebraic Topology}{Springer--Verlag}{1982}.}

\ref\CLM{S.\ Cappell, R.\ Lee and E.\ Miller,
 \preprint{Invariants of 3-manifolds From Conformal Field Theory}{preprint}.}

\ref\Cr{L.\ Crane,
   \preprint{2-D Physics and 3-D Topology}{Yale University preprint (1990)}.}

\ref\DGS{F.\ Delduc, F.\ Gieres, and S.P.\ Sorella,
 \article{Supersymmetry of the D=3 Chern--Simons Action in the Langau Gauge}
   {Phys. Lett.}{B225}{1989}{367}.}

\ref\DLPS{F.\ Delduc, C.\ Lucchesi, O.\ Piguet, and S.P.\ Sorella,
 \article{Exact Scale Invariance of the Chern-Simons Theory in the Landau
Gauge}
   {Nucl. Phys.}{B346}{1990}{313-328}.}

\ref\FG{D.\ Freed and R.\ Gomph,
 \preprint{Computer Calculations of Witten's 3-Manifold Invariant}
   {U.\ of Texas Preprint}.}

\ref\Gar{S.\ Garoufalidis,
 \preprint{Relations Among 3-manifold Invariants}{ U.\ of Chicago preprint
(1991)}.}

\ref\GGRS{S.J.\ Gates, JR., M.T.\ Grisaru, M.\ Ro\v cek, and W.\ Siegal,
  \book{Superspace}{Benjamin/Cummings}{1983}.}

\ref\GMM{{E.\ Guadagnini, M.\ Martellini, and M.\ Mintchev},
 \article{Perturbative Aspects of the Chern--Simons Field Theory}
   {Phys. Lett.}{B277}{1989}{111};
 {\it Chern--Simons Field Theory and Link Invariants},
   talk at the 13th Johns Hopkins Workshop on
  ``Knots, Topology, and Field Theory``, Florence, Italy (1989).}
%

\ref\Ho{L.\ Hormander,
 \book{Linear partial differential operators {\bf 3 }}
   {Springer--Verlag}{1985}.}

\ref\ItZ{C.\ Itzykson and J.B.\ Zuber,
 \book{Quantum Field Theory}{McGraw--Hill}{1980}.}

\ref\Je{L.\ Jeffrey,
 \book{On Some Aspects of Chern--Simons Gauge Theory}
   {Oxford University Ph.D. Thesis}{1991}.}

\ref\Jo{V.F.R.\ Jones,
 \article{Hecke algebra representations of braid groups and link polynomials}
   {Ann. Math. }{126}{1987}{335--388}.}

\ref\KM{R.\ Kirby and P.\ Melvin,
 \preprint{The 3-Manifold Invariants of Witten and Reshetikhin-Turaev}
   {Berkeley preprint (1990)}.}

\ref\RS{D.B.\ Ray, and I.M.\ Singer,
 \article{R-torsion and the Laplacian on Riemannian manifolds}
   {Adv. in Math. }{7}{1971}{145--210};
 \article{Analytic torsion}{Proc. Symp. Pure Math}{23}{1973}{167}.}

\ref\RT{N.\ Reshetikhin and V.\ Turaev,
 \article{Ribbon Graphs and Their Invariants Derived From Quantum Groups}
   {Comm. Math. Phys}{127}{1990}{1-26};
 \article{Invariants of 3-manifolds via link polynomials and quantum groups}
   {Invent. Math.}{103}{1991}{547-597}.}

\ref\Sc{A.\ Schwarz,
 \article{The Partition Function of Degenerate Quadratic Functionals
   and Ray--Singer Invariants} {Lett. Math. Phys}{2}{1978}{247}.}

\ref\Wa{K.\ Walker,
 \preprint{On Witten's 3-manifold Invariants}{preprint (1991)}.}

\ref\We{P.\ West,
 \book{Introduction ot Supersymmetry and Supergravity}{World
Scientific}{1986}.}

\ref\WiI{E.\ Witten,
 \article{Quantum field theory and the Jones polynomial}
   {Comm.  Math. Phys.}{121}{1989}{351-399}.}

\ref\WiII{E.\ Witten,
 \article{Gauge Theories, Vertex Models, and Quantum Groups}
  {Nucl. Phys. B330}{1990}{286};
 \article{Gauge Theories and Integrable Lattice Models}
  {Nucl. Phys.}{B322}{1989}{629}.}

\ref\WiIII{E.\ Witten,
  \preprint{On Quantum Gauge Theories in Two Dimensions}{IAS Preprint (1991)}.}

\pageno=1

\vskip 1in
{\sevteenrm \centerline{Chern--Simons Perturbation Theory} }
\vskip .75in
{\twelverm
\centerline{Scott Axelrod}\par
\centerline{\it Math Department, Harvard University and MIT}\par
\par
\centerline{I.M.~Singer}\par
\centerline{\it Math Department, MIT}\par
 }
\vskip 1.5in
{\hsize= 5.5 true in \hoffset= .5 true in
\noindent{\bf Abstract.}
We study the perturbation theory for three dimensional
Chern--Simons quantum field theory on a general compact three manifold
without boundary.  We show that after a simple change of
variables, the action obtained by BRS gauge fixing in the Lorentz
gauge has a superspace formulation.
The basic properties of the propagator and the Feynman rules are written
in a precise manner in the language of differential forms.
Using the explicit description of the propagator singularities, we prove
that the theory is finite.  Finally the anomalous metric
dependence of the $2$-loop partition function on the Riemannian metric
(which was introduced to define the gauge fixing) can be cancelled
by a local counterterm as in the $1$-loop case [\WiI].
In fact, the counterterm is equal to the Chern--Simons action
of the metric connection, normalized precisely as one would expect
based on the framing dependence of Witten's exact solution.
}
\vfill
\footnote{}{
This work was supported in part by the Divisions of Applied
Mathematics of the U.~S.~Department of Energy under contracts
DE-FG02-88ER25065 and 
DE-FG02-88ER25066.  
}
\vfil\eject

\def\egdiag{$1$}   
\def\dumbset{$2$}  
\def\cprop{$3$}    
\def\dblob{$4$}    
\xdef \Sintro {{1}}
\xdef \Ssuper {{2}}
\xdef \saction {Eq. 2.17}
\xdef \Sclosed {{3}}
\xdef \pspone {Eq. 3.52}
\xdef \nonpert {Eq. 3.57}
\xdef \Sfiniteness {{4}}
\xdef \thmuvf {{{$4.1$}}}
\xdef \thmfin {{{$4.2$}}}
\xdef \Sformal {{5}}
\xdef \thmlmo {{{$5.3$}}}
\xdef \thmlmt {{{$5.4$}}}
\xdef \thmtla {{{$5.5$}}}
\xdef \thmmfi {{{$5.6$}}}
\xdef \Soutlook {{6}}
\def\Dz{d^{(0)}}		
\def\dlz{\delta^{(0)}}		
\def\Zf{{Z_{\hbox{free}}}}	
\def\cg{{\Bf{g}}}		
\def\Ao{{A^{(0)}}}		
\def\TR{\op{TR}}		
\def\cs{{CS}}			
\def\G{{\Bf{G}}}		
\def\Ga{{\Gamma}}		
\def\Zhlk{{Z^{hl}_k}}		
\def\adp{{\cg}}			
\def\As{{\cA}}			
\def\tdu{{\widetilde{du}}}	
\def\Ivd{{\dot I_V^{disc}}}	
\def\bd{{bd}}			
\def\sing{{sing}}		
\def\lez{{\lim_{\epsilon \rightarrow 0}}}
\def\Zke{{Z_k^{exact}}}		

\newsec\Sintro{Introduction}

One of the most successful topological quantum field theories considered
to date has been the three dimensional Chern--Simons theory with the
inclusion of Wilson loops.
In his seminal paper [\WiI] and in subsequent work [\WiII],
E.~Witten described the exact solution for this theory and
showed that the observables lead to a broad class of invariants of
compact three manifolds with imbedded knots
(also with choices of orientations, framings and labelings),
generalizing the knot invariants defined by V.~Jones [\Jo].
Witten's starting point was a Feynman path integral formulation of
the observables.  Of course from a mathematical point of view this
starting point is purely formal.  What Witten did, however, was
perform formal manipulations of the path integral, based upon
physical insight and experience, to arrive at an ansatz for the solution.
Subsequent work by several authors [\CLM], [\KM], [\Wa], [\Cr]
have verified and made rigorous Witten's main results.
Also a theory essentially equivalent to Witten's solution of
Chern--Simons was described in [\RT].

When no links are present, the invariant for
a compact oriented $3$-manifold $M$ (taken here to have no boundary)
is given by the partition function
$$Z_k(M,G) = \int \cD A~ e^{ik\cs(A)}
,\etag\wpi$$
The basic field $A$ in \wpi\ is a connection (gauge field) with compact gauge
group $G$, and the Chern--Simons action is given by
$$\cs(A) ={1\over 4\pi}\int_M \Tr( A\wedge dA + {2\over 3} A^3 )
.\etag\last$$
Here $\Tr$ is the basic trace on the Lie algebra $\cg$ of $G$ normalized so
that the pairing $(A,B) =-\Tr(AB)$ on $\cg$ is the basic
inner product\foot{
  Normalized so that it corresponds, under the
  Chern-Weil homomorphism, to a generator of $H^4(B\tilde G,\IZ)$, where
  $\tilde G$ is the simply-connected cover of $G$.
}.
For notational simplicity, we have taken the underlying principal
bundle to be trivialized and identified the connection $A$ with
a $\cg$ valued one form on $M$.

Witten's solution follows not by ``evaluating'' the path integrals
directly but by exploiting deep connections with rational conformal
field theory.  However, the motivation behind and the intuition
for the construction is that there is some a priori definition
of the path integral which behaves as expected physically.
Unfortunately, a direct definition
of the full non-perturbative path integral seems beyond the reach of our
present-day techniques.

Here we study the {\it perturbative} formulation.
Chern--Simons perturbation theory on flat $\IR^3$
has been looked at previously by several groups of physicists.
In [\GMM], the theory up to $2$-loops was found to be finite
and to give knot invariants.
In [\BRT], [\BC], [\DGS] a
superspace formulation of the gauged fixed action was given.
Two different arguments for finiteness to all orders are
presented in [\BC] and [\DLPS],
both assuming a nice regularization scheme and
employing special symmetries of the gauge fixed action
to conclude that the $\beta$ function vanishes.
In [\BN], Dror Bar-Natan gave a rigorous treatment of
the perturbative definition of knot invariants
in $\IR^3$ up to $2$ loops,
and showed it agreed with the results expected from Witten's exact solution.

In this paper, we will allow $M$ to be an arbitrary compact, oriented,
$3$-manifold without boundary and will obtain
a succinct description of the $l$-loop contribution in the language
of differential forms which we show is finite directly.
Specifically, we perturb about a solution, $A^{(0)}$, of the
equations of motion (i.e. a flat connection).
We shall assume that $\Ao$ is isolated up to gauge transformations and
that the group of gauge transformations fixing $\Ao$ is discrete.
Equivalently, we assume that the cohomology of $\Dz$ vanishes, where
$\Dz$ is the exterior covariant derivative
twisted by $\Ao$ and acting on $\Omega^*(M;\cg)$, the space of forms
with values in the associated adjoint bundle.  The differential forms
viewpoint instructs us to sum over all particle types before integrating and
provides us with a natural point splitting of the propagator on the
diagonal.  One could say this is the regularization scheme we use.

To define the perturbative expansion, it is necessary to make a choice of gauge
fixing.  We choose BRS gauge fixing using Lorentz gauge,
which depends on a choice of Riemannian metric, $g$, on $M$.
The perturbative expansion has the form
$$ Z_k(M,A^{(0)},g) = Z_k^{sc}(M,A^{(0)},g) Z^{hl}_{k+h}(M,\Ao,g)
,\etag\last$$
where $Z_k^{sc}$ is the ``semi-classical approximation''
and $Z^{hl}_{k+h}$ is the sum of the higher order corrections.
We have included an ad hoc shift in $k$ here necessary
for agreement with Witten's exact solution.
We will explain one reason this shift is needed in \S 6, although
we can offer, at present, no derivation of it.
The role of the shift
in $k$ in perturbation theory has been the subject of much discussion
in the physics literature.  An explanation of the problem is given in [\ALR].

The expansion of $\Zhlk$ in inverse powers of $k$ is given by
$$ \Zhlk= \Sum_{V=0,2,4,...} ({-ik\over 2\pi})^{-\half V} I_V^{disc}(M,\Ao,g)
,\etag\last$$
where $I_V^{disc}$ is the sum of the contributions of all Feynman
diagrams with $V$ vertices.
Equivalently, the expansion can be written as
$$ \Zhlk= \exp\left( \sum_{l=2}^\infty ({-ik\over 2\pi})^{1-l}
  I_l^{conn}(M,A^{(0)},g)\right)
,\etag\last$$
where $I_l^{conn}$ is the sum of the contribution of the
connected Feynman diagrams with $l$ loops.

Our results are best summarized by outlining the paper.

In \S 2, we give a form of the gauge fixed action which has a simple
superspace formulation \saction, and describe the superspace Feynman rules.

In \S 3, we rewrite the superspace Feynman rules in a succinct way
using the language of differential forms.
We describe the basic propagator as a two form on $M\times M$ with
values in $\cg\otimes\cg$ and state its salient properties (PL1)-(PL6).
$I_V^{disc}$ will be written as an integral over $M^V$ of a top form
obtained from the propagator.

In \S 4, we sketch the proof of finiteness, in particular that the multiple
integral over $M^V$ defining $I_V^{disc}$
converges despite the singularity of the propagator
on the diagonal.
The logic of the proof is as follows.
By a fairly general argument, the proof reduces
to showing convergence of diagrams for the flat space theory
which may also
have insertions of edges with a propagator given by a subleading
term in the singularity of the curved space propagator.
By the convergence theorem,
it thus suffices to show that any such flat space diagrams which
are superficially divergent vanish.
The latter follows by a simple symmetry and power counting argument.

In \S 5, we give a formal proof that $I_V^{disc}(M,\Ao,g)$ is
independent of the metric $g$.  By that we mean, a proof that uses
integration by parts ignoring the singularities.  For the case
of two loops, we give a careful treatment using Stoke's theorem.
We find an explicit anomaly given as a local integral over $M$ of
the form one would expect from power counting and symmetry
considerations.  The overall coefficient of the anomaly agrees
with what one would predict from Witten's exact solution.
This means that we obtain a manifold invariant by subtracting
a concrete counterterm from $I_2^{disc}=I_2^{conn}$.

In \S 6, we make some comments about the probable relation of the results
here to Witten's exact solution and about possible extensions of
our results.

A more detailed exposition is in preparation [\AS].
In it will be found the derivation
of the singular behaviour of the propagator near the diagonal
and a careful discussion of signs and symmetry factors.


\newsec\Ssuper{Superpace Form of Gauge Fixed Action and Feynman Rules}

In this section we derive a form of the gauge fixed action which
has a superspace formulation.  In fact,
the gauge fixed action will be seen to have the same form as
the original Chern--Simons action, but applied to a superfield with
certain constraints.  We will first perform
standard BRS gauge fixing in the Lorentz gauge, and then will
change variables and integrate out the field multiplying the
gauge fixing condition.  We arrive at a form of the gauge fixed action
which may be written in superspace.

Expanding around $A^{(0)}$, the Chern--Simons action takes the form
$$ \cs(A^{(0)}+A)= \cs(A^{(0)})
  + {1\over 4\pi}\int_M \Tr(A\wedge \Dz A + {1\over 3}A\wedge[A,A])
,\etag\last$$
where $A$ is a Lie algebra valued one form.

For the standard gauge fixing, we introduce Fermionic fields
$c$ and $\bar c$ and a Bosonic field $b$, all valued in the Lie algebra of
the group of gauge transformations, $\Omega^0(M;\adp$).
The BRS operator $Q$ is given by
$$\eqalign
{ QA = - \Dz c - [A,c], \qquad & Qc=\half [c,c] \cr
  Q\bar c=b, \qquad                & Qb=0 .
}\etag\last$$
The operator $\Dz$ is the direct sum of operators
$\Dz_q:\Omega^q(M;\cg)\rightarrow \Omega^{q+1}(M;\cg)$.
To define the Lorentz gauge condition we choose a Riemannian metric $g$
on $M$.  This allows us to define the Hodge $*$ operator
$*:\Omega^q(M;\adp)\rightarrow \Omega^{3-q}(M;\adp)$ which
satisfies $*^2=1$.  We choose the sign of $*$ so that the
inner product
$$<\chi,\psi>\equiv -\int_M\Tr(*\chi\wedge\psi),
 \qquad\hbox{for }\chi,\psi\in \Omega^q(M;\adp)
\etag\last$$
is positive.  Relative to this inner product,
the adjoint of $\Dz_q$ is
$(\Dz_q)^{\dag}= (-1)^{q+1}* \Dz_{2-q}*$.
$\Dz{}^{\dag}$ will also be denoted by $\dlz$.

The Lorentz gauge condition is $\dlz A = 0$.
This condition is implemented by the gauge fixed action
$$\eqalign{
 S_{gf}(A,c,\bar c,b) &= k~\cs(A^{(0)}+A) + QV  \cr
             V        &= \alpha <\bar c, \dlz A >,
}\etag\last$$
where $\alpha$ is a constant which we are free to select.
Choosing $\alpha = k/ 2\pi$, and defining
$\cC= *\Dz\bar c$ and $\cB= * \Dz b$, we find
$$\eqalign{
 S_{gf}(A,  & c,\bar c ,b) - k\cs(A^{(0)})  \cr
  &= {k\over 2\pi}\int_M\Tr\left(\half A\wedge \Dz A +{1\over 6}A[A,A]
               - \cC\wedge \Dz c -\cC\wedge[A,c] -\cB\wedge A\right)
.}\etag\last$$
By acyclicity of $A^{(0)}$ and elementary Hodge theory,
the change of variable from $(b,\bar c)$ to $(\cB,\cC)$ is an invertible
map from pairs of elements of $\Omega^0(M;\adp)$ to pairs of
elements of $\Ker(\dlz_1)$.  Note that the Jacobian for this change of
variable is $1$ because the Fermionic determinant for the change from
$\bar c$ to $\cC$ cancels the Bosonic determinant for the change
from $b$ to $\cB$.  Also note that the integral over the field
$\cB$ just imposes the Lorentz gauge constraint $\dlz A=0$.
Thus the gauge fixed path integral can be written
$$Z_k(M,A^{(0)},g)
 = \int \cD A\cD c\cD\bar c\cD b e^{iS_{gf}(A,c,\bar c,b)}
 =e^{ik\cs(A^{(0)})}\int \cD A\cD c\cD C~ e^{-S(A,c,\cC)}
,\etag\pit$$
where the last integral is over the Bosonic field $A\in\Ker(\dlz_0)$
and the Fermionic fields $c\in\Omega^0(M;\adp)$ and
$\cC\in\Ker(\dlz_1)$, and the (imaginary) action is given by
$$ S(A,c,\cC) =-{ik\over2\pi}\int_M \half A^a\wedge\Dz A^a-\cC^a\wedge\Dz c^a
  + {1\over 6}f_{abc} (A^a \wedge A^b\wedge A^c  - 6 \cC^a\wedge A^b \wedge
c^c)
.\etag\action$$
Here we have chosen coordinates on $\cg$ corresponding to
an orthonormal basis $T_a$ (relative to the basic inner product).
The (totally antisymmetric) structure constants $f_{abc}$ of $\cg$ are
defined by
$$[T_b,T_c]= f_{abc} T_a
.\etag\last$$

\bigskip

To give \action\ a superspace interpretation,
we need some supermanifold
notation.  For $V$ a vector space, we let $V_-$ denote a Fermionic
copy of $V$.  More generally, for $E$ a vector bundle over a base
manifold $N$, we let $E_-$ denote $E$ but with the fibers considered
Fermionic.  So functions on $E_-$ are sections over $N$
of $\Lambda^*(E)$\foot{
  More precisely, Bosonic (Fermionic) functions
  on $E_-$ correspond to a choice of a Bosonic (Fermionic) section
  of $\Lambda^{even}(E)$ together with a Fermionic (Bosonic) section
  of $\Lambda^{odd}(E)$.
}.

Our base supermanifold is $TM_-$, that is
the tangent bundle to $M$,
but with the fibers considered to be Fermionic (and the base still
bosonic).  A local coordinate system $x^\mu$ on the base $M$ determine
Fermionic coordinates $\theta^\mu$ on the Fibers of $TM_-$.
The $\theta^\mu$ essentially behave like the one forms $dx^\mu$.
In fact, there is a correspondence between
differential forms, $\tilde a$,  on $M$ and functions, $a$, on $TM_-$
(i.e. ``superfields'') given by
$$ \tilde a(x)=\sum_i a_{\mu_1...\mu_i}dx^{\mu_1}...dx^{\mu_i} \leftrightarrow
   a(x,\theta)=\sum_i \theta^{\mu_1}...\theta^{\mu_i}a_{\mu_1...\mu_i}
.\etag\cor$$
Under this correspondence,
integration of top forms corresponds
to integration of superfields in the natural supervolume form on $TM_-$,
the exterior differential operator $d$ on forms
corresponds to the operator $\theta^\mu\part{x_u}$ on superfields, and
wedge product of differential forms corresponds
to multiplication of superfields.
Care must be taken with the correspondence of products because in
the forms language one takes the $dx^\mu$ to commute
with Fermions, whereas in the superspace language the $\theta^\mu$
anti-commute with Fermions\foot{
   Let $a^{(i)}=\theta^\mu_1...\theta^\mu_i a_{\mu_1...\mu_i}$  denote
   the piece of $a$ of degree $i$ in the
   $\theta^\mu$, $\tilde a^{(i)}$ denote the piece of
   $\tilde a$ which has form degree $i$,
   and $|a|=\pm 1$ denote the statistics of $a$.
   Then $|a_{\mu_1...\mu_i}|=(-1)^{|a|+i}$.
   The product of $a$ with another superfield $b$ is
   $$ a(x,\theta) b(x,\theta)=\sum_{i,j} -1^{(i+|a|)j}
     \theta^{\mu_1}...\theta^{\mu_i}\theta^{\nu_1}...\theta^{\nu_j}
     a_{\mu_1...\mu_i}(x) b_{\nu_1...\nu_j}(x)
   .\etag\last$$
   Thus
   the superfield $a b$ obtained by superfield multiplication corresponds
   to the form $\sum_{i,j} (-1)^{ij+|a|j}~\tilde a^{(i)}\wedge\tilde
   b^{(j)}$.
}.

Now let $\As$ be a Fermionic $\adp$ valued superfield on
$TM_-$.  The operators $\Dz$ and $\dlz$ on $\adp$ valued forms
correspond to operators $\Dz$ and $\dlz$ on the superfield $\As$.
So that we can make contact with \pit,
we name the component fields of $\As$ as follows,
$$ \As^a(x,\theta) = c^a(x) + \theta^\mu A^a_\mu(x)
     + \theta^\mu\theta^\nu\cC^a_{\mu\nu}(x)
     + \theta^\mu\theta^\nu\theta^\rho B^a_{\mu\nu\rho}(x)
.\etag\last$$
Since $\As$ is Fermionic,
$c$ and $\cC$ are Fermionic and $A$ and $B$ are Bosonic.
Observe that the condition $\dlz \As=0$ means precisely that
$A$ is in
$\Ker(\dlz_0)$, $\cC$ is in $\Ker(\dlz_1)$, and $B$ equals
zero.  (The last fact follows from acyclicity of $A^{(0)}$.)
So the set of $\As$ for which $\dlz\As$ vanishes is equal to
the set of triples $(c,\cC,A)$ which are integrated over in \pit.
By a simple calculation, the action defined in \action\ has a
superspace formulation:
$$ S(\As)=S(c,\cC,A)
  = \lambda\int dX~ \left[\half \As^a(X)(\Dz\As)^a(X)
        +{1\over 6} f_{abc} \As^a(X)\As^b(X)\As^c(X)\right]
.\ftag\saction$$
For convenience in writing down the Feynman rules, we have introduced
the symbol $X$ for the coordinates $(x,\theta)$,
the abbreviation $dX$ for the volume form $d^3x d^3\theta$ on $TM_-$,
and the constant
$$\lambda = -ik/ 2\pi .\etag\last$$

In summary, the path integral is given by
$$ Z_k(M)=\int_{\{\As;\dlz\As=0\} }\cD\As~ e^{-S(\As)}
,\etag\spit$$
where $\As$ is a $\adp$ valued Fermionic superfield on $TM_-$ and the
action $S$ is given by \saction.

\bigskip

To write the Feynman rules in the superspace formulation, we
must take into account two complications not usually
present in the derivation of super-Feynman rules
[\GGRS], [\We].
The first complication is to deal properly with the constraint
$\dlz \As=0$.  The second complication is to keep careful
track of the overall sign in front of each graph.  Care must be taken
here because
the basic superfield is Fermionic and because the operation of
integration over the base supermanifold $TM_-$ and the operator $\Dz$
are Fermionic.

Let
$$ \Zf[\cJ]=\int_{\As\in\Ker(\dlz)} \cD\As ~
   e^{\int dX~\left[ -{\lambda\over 2} \As^a(X)(\Dz\As)^a(X)
	+ \cJ^a(X)\As^a(X)\right]}
\etag\freej$$
be the partition function for the free theory coupled to a source
$\cJ(X)$.  $\cJ$ is taken to be a Bosonic $\adp$ valued superfield,
so that the source term in \freej\ is Bosonic.
To keep track of the
constraints we introduce the operators $\hat \pi_d$ and $\hat \pi_\delta$
on $\adp$ valued superfields which orthogonally project
onto the image of $\Dz$ and
$\dlz$, respectively.  By acyclicity of $A^{(0)}$ and Hodge theory,
we have
$$\hat\pi_d +\hat\pi_\delta = \hat \delta^\adp
,\etag\last$$
where $\hat \delta^\adp$ is the identity operator acting on $\adp$
valued superfields.  ($\hat \delta$ will denote the identity operator
on ordinary superfields.)
In order to complete the square in the exponent in \freej\ and evaluate
$\Zf[J]$ in the standard way, we introduce the Fermionic operator
$\hat L$ which is the ``Hodge theory'' inverse of $\Dz$.  To define it,
first let $\hat L_1:\Im(\Dz)\rightarrow\Ker(\Dz)^{\dag}$ be the inverse of
$\Dz$ as a map from the orthocomplement of its kernel to its image.  Then
$\hat L$ is the operator on $\adp$ valued superfields obtained by first
using $\hat\pi_d$ to orthogonally project onto
$\Im(\Dz)$ and then applying $\hat L_1$.
This definition is equivalent to the equations
$$ \eqalign{
 \Ker(\hat L) &=\Im(\Dz)^{\dag} =\Ker(\dlz) 	\cr
 \Im(\hat L)  &=\Ker(\Dz)^{\dag} =\Im(\dlz) 	\cr
 \Dz\circ\hat L &=\hat\pi_d			\cr
 \hat L\circ \Dz & =\hat\pi_\delta
.}\etag\dflh$$
Note that this definition of Hodge theory inverse makes sense even
if $A^{(0)}$ is not acyclic, useful in generalizing the
results here to allow for zero modes.
$\hat L$ can also be defined by introducing the Laplacian
$$ \nabla= \dlz\Dz+\Dz\dlz
\etag\last$$
which is invertible by
acyclicity of $A^{(0)}$.
The definition \dflh\
is equivalent to
$$\hat L =\dlz\circ\nabla^{-1}
.\etag\last$$

Having introduced $\hat L$, we may complete the square,
$$\eqalign{
\int dX~ \bigg[-{\lambda\over 2} \As^a(X) (\Dz\As)^a(X)
               + & \cJ^a(X)\As^a(X) \bigg]\cr
 =\int dX~\bigg[
  -{\lambda\over 2}(\As-(\lambda^{-1} &\hat L \cJ))^a(X)
                   \big[\Dz(\As-(\lambda^{-1}\hat L\cJ))\big]^a(X)     \cr
        & + {1\over 2} (\lambda^{-1} \hat L\cJ)^a(X) \cJ^a(X)\bigg]
,}\etag\last$$
and shift the variable $\As$ in the usual way to obtain
$$ \Zf[\cJ] = \Zf[0] e^{{1\over 2}\int dX~(\lambda^{-1}\hat L\cJ)^a(X)
\cJ^a(X)}
.\etag\freez$$

\medskip

Now we want to write $\hat L$ as an integral operator with an
integral kernel $L(X,Y)$.
For any operator $\hat K$ on $\adp$-valued superfields
(either Bosonic or Fermionic),
we define the corresponding integral kernel
$K$ to be the superfunction on $TM_-\times TM_-$ with values in
$\adp\times\adp$ so that
$$ (\hat K\Psi)_a(X) = \int dY~K_{ab}(X,Y) \Psi_b(Y)
.\etag\last$$
Note that since the operator $\int dY$ is Fermionic, the integral kernel
$K$ will have the opposite statistics to the operator $\hat K$.

The kernel $\delta$ for the identity operator $\hat\delta$ on superfields,
which one might call the ``super-delta function'',
is Fermionic and satisfies
$$\int dX dY~\delta(X,Y) \Psi(Y)\Phi(X) =\int dX~\Psi(X)\Phi(X)
.\etag\last$$
This implies more generally that
$$\int dXdY~\delta(X,Y)\Psi(X,Y,Z) =\int dX~\Psi(X,X,Z)
\etag\pdual$$
for $\Psi$ a function of three variable in $TM_-$.
Since $dYdX$ equals $-dXdY$,  the super-delta function is antisymmetric
under exchange of $X$ and $Y$.
Similarly, $\delta^\adp_{ab}(X,Y)=\delta_{ab}\delta(X,Y)$ is antisymmetric
under simultaneous exchange of $(X,a)$ and $(Y,b)$.
One can show likewise that the kernel for $\hat L$ is antisymmetric,
$$L_{ab}(X,Y)= -L_{ba}(Y,X)
,\etag\last$$
as one would expect for a Fermionic propagator.

Next we write the partition function with interaction in terms of
the free partition function with source in the usual way,
$$ Z_k = e^{i S(\Ao)}\exp \left(-{\lambda\over 3!}\int dX ~
           f_{abc}\part{\cJ^a(X)}\part{\cJ^b(X)}\part{\cJ^c(X)}\right)
  \Zf[\cJ]|_{\cJ=0}
.\etag\fone$$
The expression $\part{\cJ^a(X)}$ appearing here is defined as follows.
Given a variation
$\delta\cJ$ of $\cJ$, we let $\delta_{\delta\cJ}$ be the operator
of differentiation in the direction of $\delta\cJ$.  This acts on
functionals $\Omega(\cJ)$ of $\cJ$.  The operator
$\part{\cJ^a(X)}$ is then defined by
$$\delta_{\delta\cJ}\Omega(\cJ)
   =\int dX~\delta\cJ^a\left[\part{\cJ^a(X)}\Omega(\cJ)\right]
.\etag\last$$
So, for example, we have
$$\part{\cJ^a(X)}\int dY~\cJ^a(Y)\As^a(Y) = \As(X)
.\etag\last$$
Since $\cJ$ is Bosonic, $\part{\cJ^a(X)}$ is Fermionic, and so
$\part{\cJ^a(X)}\int dY$ equals $-\int dY\part{\cJ^a(X)}$.  Hence
$$\part{\cJ^a(X)} \cJ^b(Y) =
-\delta_{ab}\delta(X,Y)=\delta_{ba}\delta(Y,X)
.\etag\last$$

Expanding out the exponentials in \fone\ and \freez, and writing
$\hat{L}$ in terms of the kernel $L$, we find
$$Z_k= Z_k^{sc}\Sum_{I,V=0}^\infty
 {(-\lambda/3!)^V\over V!} {\lambda^{-I}\over I!}
 \left[(\Pi_{i=1}^V F_i ) \cL^I \right]|_{\cJ=0}
,\etag\ftwo$$
where
$$\eqalign{
 Z_k^{sc} &= e^{ik\cs(\Ao)} \Zf[0] \cr
 F_i &=\int dX~f_{a^i b^i c^i}~
\part{\cJ^{a^i}(X_i)}\part{\cJ^{b^i}(X_i)}\part{\cJ^{c^i}(X_i)},\cr
\cL  &= \half\int dYdZ~ \cJ^a(Y) L^{ab}(Y,Z)\cJ^b(Z)
.}\etag\last$$
We may restrict the sum in \ftwo\ to $V$ and $I$ satisfying $ 3V= 2I$
since all other terms vanish.

To write $\left[(\Pi_{i=1}^V F_i ) \cL^I \right]|_{\cJ=0}$ in a
form not involving functional derivatives, we consider sources $\cJ$
of the form
$$ \cJ^a(Y) = \Sum_{i=1}^V j^a_{(i)} \delta(Y,X_j)
.\etag\last$$
Here $j_{(i)}$ is a Fermionic $\adp$ valued source at the $i^{th}$ position.
Using
$$ \part{\cJ^a(X)}\cJ^b(Y)
  =\delta_{ab}\delta(Y,X^i) = \part{j^a_{(i)}} \cJ^b(Y)
,\etag\last$$
we find that the correction to the semiclassical approximation is
given by
$$\eqalign{ \Zhlk &\equiv { Z_k\over Z_k^{sc}} \cr
  &=\Sum_{3V=2I} {\lambda^{V-I} \over (3!)^V(2!)^I V!I!}
  \Pi_{i=1}^V\left[\int dX_i ~f_{a^ib^ic^i}
            \part{j^{a^i}_{(i)}}\part{j^{b^i}_{(i)}}\part{j^{c^i}_{(i)}}
       \right]_{j_{(i)}=0}
              L_{tot}^I
,}\etag\zthree$$
where, for given $V$,
$$\eqalign{
 L_{tot}  &= \sum_{i,j=0}^V L_s(X_i,j_{(i)},X_j,j_{(j)}) \cr
 L_s(X_i,j_{(i)},X_j,j_{(j)}) &\equiv L_{ab}(X_i,X_j) j^a_{(i)}j^b_{(j)}
.}\etag\lslt$$
For the meaning of this when $i$ equals $j$ see \pspone.
$L_s$ is the propagator $L$, but rewritten as a Bosonic
superfunction on $(TM\oplus\adp)_- \times (TM\oplus\adp)_-$.

At the level of diagrams, \zthree\ means the superspace Feynman rules assign
a factor
$$\int dX ~\left[f_{abc}
            \part{j^{a}}\part{j^{b}}\part{j^{c}}
       \right]_{j=0}
,\etag\last$$
to a vertex labeled by the point $(X,j)\in [TM\oplus\adp]_-$,
and a propagator $L_s(X,j,X',j')$ on an edge between vertices
labeled by $(X,j)$ and $(X',j')$.
These Feynman rules have a superspace formulation on $[TM\oplus\adp]_-$
even though the original action only had $TM_-$ as the base supermanifold
--the Lie algebra directions were not supercoordinates there.  In fact, the
superfield $\cA$ can be viewed as a bosonic superfield
$\cA(x,\theta,j)=\cA^a(x,\theta) j_a$
on $[TM\oplus\adp]_-$ satisfying constraints, $\dlz\cA=0$ and
$\cA$ depends linearly on $j$.
The cubic ``potential'' term
in the action \saction\
can be written as an integral over $[TM\oplus\adp]_-$,
in which $\Tr$ becomes a constant superfield.  However, the
quadratic ``kinetic'' term cannot be naturally described this way.
One sees similar phenomena in other superspace field theories;
we hope our techniques will illuminate the special properties of those
theories.

\newsec\Sclosed{A Closed Form for $l$-loop Invariants}

In this section we describe the results obtained in the previous section in
the language of differential forms rather than superspace.
We also summarize in differential forms language the basic properties
of the propagator.
Since the translation of results from the previous section into the
form language here is straight forward, we will not spell it out.
Readers not familiar with superspace and gauge fixing can
take the statements in this
section as a starting point (although we assume some notation from above).
Although we do not need it later in the paper, we will use the
properties to write the perturbation series in a rather
elegant form \nonpert.

We emphasize that the formulation of the Feynman rules given below
could have been derived by direct
formal manipulations of the path integral (i.e. manipulations which
have rigorous analogues for integrals over finite dimensional spaces) without
having introduced Fermionic coordinates or even BRS gauge fixing.  Such
a direct derivation, however, would not explain the simple form of the
final answer we derive naturally here.

\subsubsec{Properties of the Propagator}

First we describe the basic properties of the propagator $L$.
It is the kernel for the operator $\hat L=\dlz\nabla^{-1}$ acting on
$\Omega^*(M; \adp)$, where $\nabla$ is Laplacian associated to $\Dz$.
This means that $L$ is a section of
$\Omega^*(M\times M;\adp\otimes\adp)$,
which satisfies\foot{
   The unusual sign conventions used in this equation are discussed below.
}
$$ (\hat L \psi)^a(x) \equiv \int_{M_y} L_{ab}(x,y)\wedge\psi^b(y)
.\etag\kerdef$$
Note that this definition of $L$ uses wedge products and
integration of forms rather than inner products and integration of
functions with respect to the Riemannian volume.  This will enable
us to make metric independence as manifest as possible.
The expression $M_y$ on the right hand side of \kerdef\ is simply an
abbreviation to say that we are to integrate over the copy of $M$ paramaterized
by the $y$ variable.  Since $M$ is $3$ dimensional and $\hat L$ decreases form
degree by $1$, \kerdef\ implies $L$ is a $3-1=2$ form.

When dealing with products of several copies of $M$
throughout the rest of the paper, we will often adopt the notation
used above of
distinguishing a particular copy of $M$, and objects associated with it,
by adding the name of a variable parameterizing that copy as a subscript.
So, for example, we write
$$ L\in\Omega^2(M_x\times M_y; \adp_x\otimes \adp_y)
.\etag\listwo$$

\def\pl#1{(PL#1)}  

$L$ is equivalently defined by
$$\dlz_x L(x,y)=0,\qquad \dlz_y L(x,y)=0
\leqno\pl{0}$$
together with
$$ \Dz_{M\times M} L_{ab}(x,y)
  =(\Dz_x+\Dz_y) L_{ab}(x,y) =-\delta^\cg_{ab}(x,y)\equiv -\delta_{ab}
\delta(x,y)
.\leqno\pl{1}$$
Here $\Dz_{M\times M}$ is the exterior derivative operator on
$\Omega^*(M\times M;\adp\times\adp)$ determined by the connection $(\Ao,\Ao)$;
and $\delta$ is the Poincare dual form to the diagonal, defined by
$$\int_{M_x\times M_y} \delta(x,y)\psi(x,y)=\int_{M_x} \psi(x,x)
\qquad\hbox{for }\psi\in\Omega^*(M\times M)
.\etag\last$$

The fact that $\Dz{}^2=0$ implies $\hat L\circ \hat L=0$.  Written in
terms of the kernel $L$, this becomes
$$\int_{M_y} L_{ab}(x,y) L_{bc}(y,z) =0
.\leqno\pl{2}$$

Since it is a Fermionic propagator,
$L$ is antisymmetric under the involution of $\ad(P)\times\ad(P)$
gotten by exchanging the two copies of $\ad(P)$,
$$ L_{ab}(x,y) = -L_{ba}(y,x)
.\leqno\pl{3}$$
This can be shown directly from the fact that the involution reverses the
orientation of the base $M\times M$, which
implies that $\delta(x,y)$ is antisymmetric under exchange of $x$ and $y$.
Then, since the involution leaves the operator $\Dz_{M\times M}$ in \pl{1}\
invariant, $L$ must be antisymmetric.

Now we wish to consider how $L$ varies as the metric $g$ changes by
an infinitesimal variation $\delta g$.  For $K$ an object depending
on $g$, we will denote the derivative of $K$ in the direction
$\delta g$ by either $\dot K$ or $\delta_{\delta g} K$.
First notice that properties
\pl{1} -\pl{3}\ can be stated without reference to the metric $g$.
In particular, the right hand side of \pl{1}\ does not vary as one
changes the metric.  Therefore $\Dz_{M\times M}\dot L$ vanishes.
But acyclicity of $\Dz$ implies acyclicity of $\Dz_{M\times M}$.
So the fact that $\dot L$ is closed implies it is exact,
$$ \dot L\equiv \delta_{\delta g} L = \Dz_{M\times M} B
.\leqno\pl{4}$$
Here $B$ is a $\adp\times\adp$ valued one form on $M\times M$ which is
also a one form on the space of metrics (i.e. it depends linearly
on $\delta g$).

The next important property of the propagator $L$ is the explicit
description of its singularities and discontinuities near the
diagonal, which we now describe.  For $(x,y)$ in a neighborhood
of the diagonal, using the Hadamard parametrix method [\Ho] we
find\foot{
   In our conventions, $R_{\alpha\beta\rho\sigma}=
   (R_{\alpha\beta})^\tau{}_\sigma g_{\rho\tau}$, where
   $(R_{\alpha\beta})^\tau{}_\sigma dz^\alpha dz^\beta$
   is the curvature tensor thought of as an $End(TM)$ valued $2$-form.
}
$$\eqalign{
L(x,y) &= L^{had}(x,y) + L^{cont}(x,y)  \cr
L^{had}(x,y) &= L^{sing}(x,y) + L^{\bd}(x,y)               \cr
  4\pi L^{sing}_{ab}(x,y) &= -\half \det(g)^\half
      \epsilon_{\mu\nu\rho} {u^\mu\over ||u||^3} \tdu^\nu \tdu^\rho
      \delta_{ab} \cr
  4\pi L^{\bd}_{ab}(x,y) &= -g_{\mu\nu} {u^\mu\over ||u||} \hat\cR^\nu
      \delta_{ab}  \cr
  \hat\cR^\nu &= \half\det(g)^{-\half}\epsilon^{\nu\rho\sigma}
                  R_{\alpha\beta\rho\sigma} dz^\alpha dz^\beta
,}\leqno\pl{5}$$
where $L^{cont}$ is smooth away from the diagonal and continuous
across the diagonal.
Here $z\in M$ and $u\in T_z M$ are related
to $x$ and $y$ by the exponential map,
$$(x,y) =\exp{}_{(z,z)}((u,-u))
.\etag\expmap$$
The ``horizontal'' one forms $\tdu^i$ are defined by
$$ \tdu^\mu = du^\mu + \Gamma^\mu_{\nu\rho} u^\nu dx^\rho
.\etag\last$$
The quantity $||u||$ is the Riemannian norm of $u$.
In the third and fourth lines of \pl{5},
we have identified the Lie algebra at $x$ with the Lie
algebra at $y$ using the parallel transport operator determined by $\Ao$
along the short geodesic from $x$ to $y$ (of length $2||u||$).
In [\AS], we derive \pl{5}\ from the Hadamard construction
of the kernel for $\nabla^{-1}$, and also give an alternate derivation
using equivariant differential forms.

Note that $L^\sing$ diverges quadratically as one
approaches the diagonal and $L^\bd$ is discontinuous but bounded.

For the proof of finiteness in \S 4, it will be also be useful to
describe the propagator singularities in the following way
(which, with some care, can be derived from the above).
For $z^{(0)}\in M$ and small $y_1,y_2\in T_{z^{(0)}} M$, we let
$x_i =\exp_{z^{(0)}}(y_i)$, $i=1,2$.  Then
$$ L_{ab}(x_1,x_2) = -\delta_{ab} \left[
  \epsilon_{\mu\nu\rho} \hat u^\mu d\hat u^\nu d\hat u^\rho~\alpha
 +\epsilon_{\mu\nu\rho} \hat u^\mu d\hat u^\nu \gamma^\rho\right]
 +(bounded)_{ab}
,\etag\regname$$
where $u=y_1-y_2$, $\hat u= u/||u||$,
$\alpha$ is a bounded function of $y_1$ and $y_2$,
$\gamma^\rho$ is a bounded $1$-form for each $\rho$,
and $(bounded)_{ab}$ is a bounded $2$-form for each $a$ and $b$.

One can also derive an explicit description of the singularities of
$B$ near the diagonal\foot{
  Here $\Gamma_{\alpha\rho\sigma}
     = g_{\sigma\tau} (\Gamma_\alpha)^\tau{}_\sigma$, where
  $(\Gamma_\alpha)^\tau{}_\sigma dz^\alpha$
  is the connection one-form for the metric connection.
},
$$\eqalign{
B(x,y) &= B^{had}(x,y) + B^{cont}(x,y)  \cr
B^{had}(x,y) &= B^{sing}(x,y) + B^{\bd}(x,y)               \cr
  4\pi B^{sing}_{ab}(x,y) &=
   \half \det(g)^\half\epsilon_{\mu\nu\rho} {u^\mu\over ||u||^3}\tdu^\nu
         (g^{\rho\sigma}\delta g_{\sigma\tau} u^\tau)\delta_{ab} \cr
  4\pi B^{\bd}_{ab}(x,y) &=
   g_{\mu\nu} {u^\mu\over ||u||} \hat\cO^\nu	\delta_{ab}	 \cr
  \hat\cO^\nu &= \half\det(g)^{-\half}\epsilon^{\nu\rho\sigma}
  \cO_{\alpha\rho\sigma} dz^\alpha \cr
  \cO_{\alpha\rho\sigma} & =
   (\delta\Gamma_{\alpha\rho\sigma} -\half \nabla_\alpha\delta g_{\rho\sigma})
.}\leqno\pl{6}$$
Note that $B^\sing$ diverges linearly as one approaches the diagonal,
and $B^\bd$ is bounded and discontinuous.

\subsubsec{Form of the Perturbative Expansion}

Having stated the important properties of the propagator in terms of
forms, we will state the precise transcription of \zthree\ and \lslt\
into that language.

Let
$$\Gamma_{x_1...x_V}
   =\Gamma(M_{x_1}\times ... \times M_{x_V},
     \Lambda^*(\oplus_{i=1}^V([T^*M_{x_i}\oplus\adp_{x_i}]) ))
\etag\last$$
denote the space of sections of the bundle over
$M_{x_1}\times... \times M_{x_V}$ whose fiber at $(x_1,...,x_V)$
is the graded Grassmann algebra generated by one forms
$ dx_i^\mu$ and $j_{x_i}^a$, $i=1,...,V$,  corresponding to bases elements for
the cotangent space of $M$ and the adjoint bundle of $P$
at the points $x_i$.
Multiplication on $\Gamma_{x_1...x_V}$ is pointwise wedge product.
The one forms $j_{x_i}^a$ will also sometimes be denoted by $j_{(i)}^a$.
The operation of interior product with $j_{(i)}^a$ will be denoted
$\part{j_{(i)}^a}$\foot{The definition of this requires a trivialization
of $\adp$, and so can only be defined locally when $\adp$ is
nontrivial.  But the equations below are always valid globally.
}.

Given an element $A\in\Omega^*(M_x\times M_y;\adp_x\otimes\adp_y)$
(e.g. the propagator $L$), we define $A_s\in\Gamma_{xy}$ to be the image of $A$
under the natural injection from
$\Omega^*(M_x\times M_y; \adp_x\otimes\adp_y)$ to
$$\Gamma_{xy}=\Gamma(M_x\times
M_y;\Lambda^*([TM_x\oplus\adp_x]\oplus[TM_y\oplus\adp_y]))
.\etag\last$$
Also define $A_{tot}\in\Gamma_{x_1,...,x_V}$ by
$$A_{tot}(x_1,...,x_V) = \Sum_{i,j=1}^V A_s(x_i,x_j)
.\etag\dtp$$
Explicitly, the $A_s(x_i,x_j)$ appearing here are given by
$$ A_s(x_i,x_j) = A_{ab}(x_i,x_j) j_{(i)}^a j_{(j)}^b
.\etag\last$$

Although the propagator $L$ is singular along the diagonal,
the singularity is symmetric in the group theory indices, because it
is proportional to $\delta_{ab}$.  Hence
$L_s$ extends continuously across the diagonal,
$$ L_s(x_i,x_i)\equiv L^{cont}_{ab}(x_i,x_i) j_{(i)}^a j_{(i)}^b
.\ftag\pspone$$
(This regularization can be stated in any number of equivalent
ways as a point splitting regularization.)

Before going on, we make a rather technical comment about our sign
conventions in \kerdef\ which is a translation into forms language
of conventions built into the superspace language.
We adopt the unusual sign convention that the expression
$\int_{M_y}$ in \kerdef\ is defined by
$$\int_{M_y} [\psi(y)\chi(x)] \equiv [\int_{M_y}\psi(y)]\chi(x)
\qquad\hbox{for }\chi\in\Omega^*(M_x),\psi\in\Omega^*(M_y)
.\etag\nonstd$$
This convention is opposite to the usual mathematical definition
(see e.g. [\BT], p. 61]) and means that the exterior
derivative operator anticommutes with $\int_{M_y}$ rather than
commuting.  This corresponds, in the notation of the previous section,
to the fact that $d\int dX F(X,Y)=-\int dX~dF(Y,X)$.
More generally, we define the operator $\int_{M_{x_i}}$ on
$\Gamma_{x_1...x_V}$ by stating that
for $\psi(x_i)$ in $\Gamma_i$ and $\chi(x_1,...,x_{i-1},x_{i+1},...,x_V)$
in $\Gamma_{x_1...x_{i-1},x_{i+1},...,x_V}$,
$$\int_{M_{x_{i_1}}} [\psi(x_i) \chi(x_1,...,x_{i-1},x_{i+1},...,x_V)]$$
equals
$$[\int_{M_{x_i}} \psi(x_i)] \chi(x_1,...,x_{i-1},x_{i+1},...,x_V)$$
if $\psi$ sits inside $\Omega^*(M)$
and equals zero if $\psi$ is of degree one or more in the $j_{(i)}^a$.
Note that, because we chose the nonstandard convention  in \nonstd,
in order for the operators $\int_{M_{x_1}\times ...\times M_{x_V}}$ and
$\int_{M_{x_1}}...\int_{M_{x_V}}$ to agree, we must equip
$M^V=M_{x_1}\times ...\times M_{x_V}$ with the nonstandard orientation
(for which $\mu_V...\mu_1$ is a positive volume form on the product
if the $\mu_i$ are positive volume forms on the factors).

Now \zthree\ becomes
$$\eqalign{ \Zhlk &\equiv { Z_k\over Z_k^{sc}} \cr
  &=\Sum_{3V=2I} {(-ik/2\pi)^{V-I} \over (3!)^V (2!)^I V!I!}
  \Pi_{i=1}^V\left[\int_{M_{x_i}} f_{a^ib^ic^i}
            \part{j^{a^i}_{(i)}}\part{j^{b^i}_{(i)}}\part{j^{c^i}_{(i)}}
	     \right]
              L_{tot}(x_1,..,x_v)^I
.}\etag\zfour$$

Let $\Tr_{(i)}$ be the operator
$f_{a^ib^ic^i}
 \part{j^{a^i}_{(i)}}\part{j^{b^i}_{(i)}}\part{j^{c^i}_{(i)}}$,
and let $\TR$
$$\TR: \Gamma_{x^1,...,x^V}\rightarrow \Omega^*(M_{x_1},...,M_{x_V})$$
be the composition of the operator $\Tr_{(V)}...\Tr_{(1)}$ followed
by the restriction of an element in $\Gamma_{x^1...x^V}$ to the piece
of degree $0$ in the $j^{a}_{(i)}$\foot{
   The map $\TR$ encodes the proper sign for the interaction vertices.
   It is possible to interpret it as a generalized trace.
}.
Then we can rewrite \zfour\ as
$$\Zhlk = \Sum_{3V=2I} {(-ik/2\pi)^{V-I} \over (3!)^V (2!)^I V!I!}
          \int_{M^V} \TR( L_{tot}(x_1,..,x_v)^I)
.\etag\zfive$$

Although we shall not use it anywhere else in the paper, we
rewrite $\Zhlk$ in one more way that may be useful in trying to sum
the perturbation series.
Let
$$e^M=\union_{V=0}^\infty M^V/S_V
,\etag\last$$
where $S_V$ is the permutation group of order $V$ acting
on $M^V$ by exchanging the different copies of $M$.
($e^M$ can be identified with the set of finite subsets of $M$.)
Then
$$\Zhlk = \int_{e^M} \TR\left( e^{\gamma L_{tot}}\right)
,\ftag\nonpert$$
where $\gamma = \half ({-ik/2\pi})^{-{1\over 3}} (3!)^{-{2\over 3}}$.

\subsubsec{Diagramatic Description}

We now describe \zfour\ in the language of Feynman diagrams.
For the remainder of the paper, a Feynman diagram will mean a
graph $\G$, all of whose vertices have valency $1$, $2$, or $3$.
For a diagram $\G$, $V_r(\G)$
will denote the number of $r$-valent vertices,
$V(\G)=V_1(\G)+V_2(\G) +V_3(\G)$ the total number of
vertices, $I(\G)$ the number of edges,
$C(\G)$ the number of connected components,
and $l(\G)$ the number of loops (the dimension of the first homology
group).
So we have,
$$\eqalign{
 2I(\G) &= 3V_3(\G) + 2V_2(\G) + V_1(\G),\hbox{ and}		\cr
 V(\G)-I(\G) &= C(\G)-l(\G)
.}\etag\last$$
When the diagram is clear, we will simply write $V$ for $V(\G)$,
$I$ for $I(\G)$, etc.

The diagrams we look at should be thought of as diagrams for
truncated Greens functions.  The number of external legs is
defined to be
$$E(\G) = 3V(\G)-2 V(\G) = V_2(G) + 2V_1(G)
.\etag\last$$
For example, in Figure \egdiag\ the external legs are
numbered from $1$ to $5$.

We now write down a generalization of the amplitude for
a truncated graph $\G$ coupled to sources at the external legs.
To do so, we order the vertices and let $x_i$ be the name of
a variable in $M$ labeling the $i$'th vertex.  The generalization
of a collection of external sources
will be an element $\Psi$ of $\Gamma_{x_1,...,x_n}$.
We let $\cI(\G)$ be the product of the propagators for each of the
edges of $\G$.  To write this down explicitly,
we choose an orientation of the graph and an ordering of the edges.
Let $in(e)$ ($out(e)$) be the incoming (outgoing) vertex of the $e$'th edge.
Then $\cI(\G)$ is given by
$$\cI(\G)(x_1,...,x_V) = \Pi_{e=1}^I L_s(x_{in(e)},x_{out(e)})
.\etag\last$$
The amplitude for the graph $\G$ coupled to the source $\Psi$ is
$$\int_{M_{x_1}\times ...\times M_{x_V}}
  \TR(\cI(\G)(x_1,...,x_V) \Psi(x_1,...,x_V))
;\etag\amppsi$$
or, more succinctly, $\int_{M^V}\TR(\cI(\G)\Psi)$.

As an example, to get the amplitude for Figure \egdiag\ with a source
$\cJ_i$ flowing in at the $i$'th external leg, we take
$$\Psi(x_1,x_2,x_3,x_4,x_5)=
\cJ_1(x_1)\cJ_2(x_1)\cJ_3(x_4)\cJ_4(x_4)\cJ_5(x_5)
.\etag\last$$

The Feynman amplitude for a diagram
$\G$ with no external legs (i.e. a trivalent graph), is the amplitude
when $\Psi$ is equal to $1$,
$$I(\G) \equiv \int_{M^V}\TR(\cI(\G))
.\etag\ig$$
For the special case of the empty graph, we set $I(\G)=1$.

In order to write $I(\G)$ in a way making the group theory indices explicit,
we make the following definition.
A {\it labeling} of a graph $\G$ is a choice of
(i) an ordering of the vertices from $1$ to $V$,
(ii) an ordering of the edges from $1$ to $I$,
(iii) an orientation, and (iv) an
ordering of the three edges incident on any given vertex.
Graphs with
labels will be denoted $\bar\G$, where $\G$ is the underlying
unlabeled graph.  Given a labeled graph $\bar\G$, one can define
an injection
$$ F:\{1,...,I\}\times\{1,2\}\rightarrow \{1,..., 3V\}
\etag\last$$
by setting $F(e,1)=3(in(e)-1)+j_{in}(e)$ and
$F(e,2)=3(out(e)-1)+j_{out}(e)$
when the $e$'th edge points from the vertex $in(e)$ to the vertex
$out(e)$ and is ordered as the $j_{in}(e)$'th edge incident on
$in(e)$
and the $j_{out}(e)$'th edge incident on $out(e)$.
Similarly, such
an injection $F$ determines a labeled graph $\bar\G$.
The map $F$ is onto precisely when the underlying graph $\G$ is
closed (has no external legs).
It will be  convenient to abbreviate $F(e,1)$ by $e(1)$, $F(e,2)$ by $e(2)$,

The Feynman amplitude for a closed, labeled diagram $\bar\G$ is
$$ I(\bar\G)  =\int_{(x_1,...,x_V)\in M^V}
     \sigma(\bar\G) f_{a^1a^2a^3}...f_{a^{3V-2}a^{3V-1}a^V}
     \wedge_{e=1}^I L_{a_{e(1)}a_{e(2)}}(x_{in(e)}, x_{out(e)})
,\etag\fbg$$
where $\sigma(\bar\G)$ is equal to $\pm 1$.
The overall sign $\sigma(\bar\G)$ is such that $I(\bar\G)$
is equal to the Feynman amplitude $I(\G)$ defined above.
This overall sign is irrelevant for the proof given in the next
section that the integral in \fbg\ is
convergent despite the singularities near the various diagonals.
For the proof of metric
independence up to local anomalies, however, the relative signs
between graphs are crucial.


The higher loop perturbation series $\Zhlk$ which is the focus of
our study is given by a weighted sum of the Feynman amplitudes of
the labeled trivalent graphs,
$$ \Zhlk = \Sum_{\bar \G} {(-ik/2\pi)^{V-I}\over (3!)^V(2!)^I V! I!}
		          I(\bar \G)
.\etag\last$$
$\Zhlk$ can be rewritten
$$ \Zhlk =\Sum_{V=0,2,...}^\infty (-ik/2\pi)^{-\half V} I_V^{disc}
,\etag\disc$$
where $I_V^{disc}$ is the contribution of all graphs (connected or not)
with $V$ vertices,
$$ I_V^{disc} =\Sum_{{\bar \G\atop V(\G)=V}}
               {1\over (3!)^V(2!)^I V! I!} I(\bar\G)
.\etag\last$$
This can also be written as a sum over unlabeled graphs,
$$ I_{V}^{disc} =\Sum_{{\G\atop V(\G)=V}}
		{1\over S(\G)} I(\G)
,\etag\last$$
where $S(\G)$ is the symmetry factor of the graph $\G$\foot{
   To define $S(\G)$, we let $P_{V,I}$ be the group of order
   $(3!)^V (2!)^I V! I!$ generated by changes of orientation on
   any of the $I$ edges, and permutations of the set
   $\{1,...V\}$ of vertex labels, the set $\{1,...,I\}$ of edge labels,
   and of the orderings of incident vertices on any of the $V$ edges.
   $P_{V,I}$ acts on the set of labeled graphs with $V$ vertices and
   $I$ edges.  The orbits of the action being sets of labeled graphs
   with the same underlying unlabeled  graph.  $S(\G)$ is the number
   of elements of $P_{V,I}$ which fix a labeled graph $\bar \G$ with
   underlying graph $\G$.
   }.

Letting $I_l^{conn}$ be the contribution of $l$-loop
connected graphs, we have
$$\Zhlk = \exp\left(
\Sum_{l=0}^\infty ({-ik\over 2\pi})^{1-l} I_l^{conn} \right)
.\etag\zcon$$

\newsec\Sfiniteness{Proof of Finiteness}

In this section we sketch the proof of finiteness of
Chern--Simons perturbation theory for all $M$, $\Ao$, and $g$ as above.
Further details will appear in [\AS].
The following strong definition of finiteness
implies that the Feynman integrals computing
all correlation functions (including the partition function itself)
are absolutely convergent.
We say that the theory is finite if, for every graph $G$ with
vertices labeled by points $x_1,...,x_V$ and every bounded section
$\Psi\in \Gamma_{x_1,x_2,...,x_V}$, the integral
$\int_{M^V} \TR(\cI(\G)\Psi)$ is convergent.

Although we have been considering theories
formulated on compact $M$,
it also makes sense to formulate
perturbation theory for noncompact manifolds.  To do so, one should
impose conditions at infinity to arrive at an
appropriate definition of the propagator, and then use the same
formulas as above for the perturbation theory.  In particular,
for flat $\IR^3$ with the trivial connection,
the propagator is the free propagator,
$$ 4\pi L^{free}_{ab}(x,y) =
   -\half \delta_{ab}\epsilon_{\mu\nu\rho} {u^\mu\over ||u||^3} du^\nu du^\rho
,\etag\freeprop$$
where $x,y\in\IR^3$ and $u$ equals $x-y$.
We say a theory on a noncompact manifold is ultraviolet finite if the
integral $\int_{M^V}\TR(\cI(\G)\Psi)$ is always locally integrable.
(For a compact manifold ultraviolet finiteness is the same as
finiteness.)

\medskip

As a warmup to the general proof of finiteness, we first prove:

\proclaim{Theorem \newthm\thmuvf}.
The theory is ultra-violet finite in flat $\IR^3$.

Then we shall prove:
\proclaim{Theorem \newthm\thmfin}.
The theory is
finite for a general oriented compact 3-manifold $M$ without boundary.

\subsubsec{Sketch of Proof of Theorem \thmuvf.}

For a general flat space theory,
the superficial degree of divergence, $\Delta(\G)$,
for a diagram $\G$ is
the degree by which one would expect its flat space Feynman integral to
diverge due to the singularities when all the vertices approach one another.
For the theory we are considering, the superficial degree of
divergence of $\G$ is
$$\Delta(\G) = 2I - 3(V-C) = 3C - E
.\etag\powcnt$$
That is, one counts plus 2 for each edge
because of the $1/||u||^2$ divergence of the propagator,
minus 3 for the integration at each vertex,
and plus 3 since the overall translation collective coordinate
for each connected component does not help with convergence.
The right hand side of \powcnt\ also equals $3l-I$ as one
one would expect from momentum space power counting:
plus 3 for the 3 momentum integrated over for each loop and
minus 1 because the momentum space propagator falls off as one power
of the momentum (since it is the kernel of a differential operator of order 1).

We call a connected diagram superficially divergent if
its superficial degree of divergence is non-negative and if
it has at least one loop (tree graphs should not be considered divergent).
A general diagram is called superficially divergent if any of its connected
subdiagrams are.

Recall the convergence theorem for Feynman integrals
which says that if a Feynman diagram has no superficially divergent
connected subdiagrams then it is absolutely convergent locally\foot{
  See [\ItZ], \S 8.1.4 and literature cited therein.
  We use a version of the theorem easily proved by some slight
  modifications of the proof described in [\ItZ].
  The proof there is for four dimensional scalar theories,
  but easily generalizes to any number of dimensions and any type of particles.
  The proof in [\ItZ] also assumes a massive theory, but that
  is only to avoid infrared divergences, which do not concern us
  here (we only need convergence locally).  Finally, the theorem in
  [\ItZ] only refers to one-particle irreducible diagrams, but that
  restriction is easily removed once one defines superficially divergent
  graphs in general as above.
}.

Thus, to prove Theorem \thmuvf,
it suffices to show that $\cI(\G)$ vanishes whenever $\G$ is a
connected superficially divergent diagram.  So let $\G$ be such a
diagram, with the vertices labeled by $x_1,...,x_V$.

Plugging $C=1$ into \powcnt,
$$\Delta(\G)= 3-E
.\etag\last$$
Combining this with $l>0$ and some algebra (or pictures), we find
that either $V=1$ and $E=1$ or $V>1$ and $E<4$.
The case $V=1$, $E=1$ is trivial because $\cI(\G)$ is $L^{free}_s(x_1,x_1)$,
which is zero in flat space.
So we may assume $V>1$ and $E<4$.

Now let
$$v^{(0)}=\Sum_{i=1}^V x_i^\mu \part{x_i^\mu}
,\etag\last$$
and
$$ v^{(\mu)}=\Sum_{i=1}^V \part{x_i^\mu}\qquad\hbox{for }\mu=1,2,3
\etag\last$$
be the vector fields on $(\IR^3)^V$ generating an overall dilation
and overall translations.
Note that, the $v^{(\alpha)}$, $\alpha=0,1,2,3$ are
linearly independent as long as the $x_i$ are not all equal.

A direct computation shows that interior product with any of these
vector fields annihilates the propagator,
$$ i(v^{(\alpha)}) L^{free}_s(x_i,x_j) = 0 \qquad\quad\hbox{for }\alpha=0,1,2,3
.\etag\last$$
Since $\cI(\G)$
is a product of propagators between different points,
it is also annihilated by interior product with any of the $v^{(\alpha)}$.
However, the form $\cI(\G)$ has degree $2I=3V-E$.
Thus $\cI(\G)$ is a form on $(\IR^3)^V$ of codimension $E$ less than $4$
which is annihilated by interior product with four vector
fields on $(\IR^3)^V$, linearly dependent almost
everywhere.  Hence $\cI(\G)$ vanishes.

\subsubsec{Sketch of Proof of Theorem \thmfin.}

Since $M^V$ is compact, it suffices to show that every
$(x_1^{(0)},...,x_V^{(0)})\in M^V$ has an open neighborhood
$U$ so that the integral
$$ I_U\equiv \int_U \TR(\cI(\G)\Psi)
\etag\last$$
is convergent for every bounded
$\Psi\in \Gamma_{x_1,...,x_V}$.
We will take $U$ to be of the form
$$U = \{ (x_1,..., x_V)\in M^V; x_i= \exp_{x_i^{(0)}}(y_i),
			y_i\in T_{x_i^{(0)}} M, ||y_i|| <\epsilon \}
,$$
where $\epsilon$ will be chosen sufficiently small.
For the rest of the proof, $(x_1,...,x_V)$ will always be a point in
$U$.

If the $x_i^{(0)}$ are all distinct, then $I_U$ converges because
the propagators are bounded in $U$.  Let $z_1^{(0)},..., z_K^{(0)}$
be the distinct points in the set $\{ x_1^{(0)},...,x_V^{(0)}\}$.
By choosing $\epsilon$ small enough, there is some constant $C$ so
that the distance between
$x_i$ and $x_j$ is greater than $C$ unless
$x_i^{(0)}$ equals $x_j^{(0)}$.
For $J$ between $1$ and $K$, let $\cI_J$ be the product of propagators
for edges $e$ connecting vertices close to $z_J^{(0)}$, i.e.
$$ \cI_J = \Pi_{e\in S_J} L(x_{in(e)},x_{out(e)}),\qquad
S_J=\{e; x^{(0)}_{in(e)}=x^{(0)}_{out(e)}= z^{(0)}_J \}
.\etag\last$$
Then
$$ I_U = \int_U \TR\left(\left[\Pi_{J=1}^K \cI_J\right] \Psi'\right)
,\etag\last$$
where $\Psi'\in\Gamma_{x_1,...,x_V}$ is bounded on $U$.
Now $U= U_1\times ... \times U_K $,
where $U_J$ is the set of positions of the $x_i$ for $i$ such that
$x_i^{(0)}= z^{(0)}_J$.
Also $\Psi'$ can be uniformly approximated by a sum of terms of the
form $\Pi_{J=1}^K \Psi_J$, where the $\Psi_J$ are bounded and
(as forms) only depend on the variables in $U_J$.
Consequently, to prove that $I_U$ converges, it suffices to show
that $ \int_{U_J}\TR(\cI(\G_J)\Psi_J) $ is finite, where $\G_J$
is the subgraph of $\G$ consisting of the vertices close to
$z^{(0)}_J$ and the edges connecting two such vertices
(so $\cI_J$ above equals $\cI(G_J)$).
Replacing $G$ above by $G_J$, we can assume that $K=1$.

To recapitulate, it suffices to show convergence of
$$I_U= \int_{{(x_1,...,x_V)\in M^V; x_i=\exp_{z^{(0)}}(y_i),
         \atop y_i\in T_{z^{(0)}} M, ||y_i||<\epsilon}}
        \TR\left(\Psi\Pi_{e=1}^E L(x_{in(e)},x_{out(e)})\right)
,\etag\irec$$
for any graph $\G$ and any bounded $\Psi$.
Now use the exponential map to pull back the integral in \irec\ to an
integral over the $y_j$'s; choose a basis $\{e_\mu\}$ of $T_{z^{(0)}} M$;
and substitute the expression in \regname.  We obtain
$$ I_U =\int_{{(y_1,...,y_V);\atop y_i\in T_{z^{(0)}}M, ||y_i||<\epsilon}}
 \TR\left(\Psi \Pi_{e=1}^I \left[
  B^e_{\mu\nu} d(\hat u_e)^\mu d(\hat u_e)^\nu +C^e_{\mu} d(\hat u_e)^\mu+D^e
  \right]\right)
,\etag\illl$$
where
$$ \hat u_e = {y_{in(e)}- y_{out(e)} \over ||y_{in(e)}- y_{out(e)}||}
\etag\last$$
and the $B$'s, $C$'s, and $D$'s are all bounded. 
The right hand side of \illl\ is a sum of terms of the form
$ \int P\wedge\omega$,
where $\omega$ is a bounded form and $P$ is a product of
the $d(\hat u_e)^\mu$.

We can think of $P$ as the Feynman integrand, $\cI(\Ga)$,  for a graph
$\Ga$ in a theory with
three different types of propagators which can be attached
to an edge $e$ connecting $y_{in(e)}$ to $y_{out(e)}$, namely the
$ d(\hat u_e)^\mu$ for $\mu=1,2,3$,
The vertex interaction is wedge product followed by
integration of top forms.
The graph $\Ga$ may have vertices of valency greater than three
and is not allowed to have any edges connecting a vertex to itself.

By the convergence theorem, it suffices to
prove that if $\Ga$ is a superficially divergent connected diagrams then
its Feynman integrand $\cI(\Ga)$ vanishes.
Now, the degree of divergence of any of the propagators $d(\hat u_e)^\mu$
is one, as is its form degree.
So the degree of divergence of $\Ga$
is $E(\Ga)- 3(V(\Ga)-1)$.  Thus $\Ga$ is superficially divergent if
$3V(\Ga) - E(\Ga)\le 3$ (and it has at least one loop).
But $E(\Ga)$ is also the form degree of $\cI(\Ga)$.
If $\Ga$ is superficially divergent,
$\cI(\Ga)$ is a form of codimension less than $4$.  However,
$\cI(\Ga)$ is annihilated by interior product with the four vector
fields generating overall dilations and translations (which are
linearly independent when $V(\Ga)>1$ as is the case for a diagram
with at least one loop and no tadpoles).
Thus $\cI(\Ga)$ vanishes if $\Ga$ is superficially divergent.

\newsec\Sformal{Formal Proof of Metric Independence and 2-loop Anomalies}

In this section we discuss the dependence of $I_V^{disc}(M,\Ao,g)$
on the metric $g$.  We give a formal proof that $I_V^{disc}$
is independent of $g$; i.e., we show that the derivative
$\delta_{\delta g} I_V^{disc}(M,\Ao,g)$ vanishes for an arbitrary
variation $\delta g$ of the metric $g$.  Our argument is formal;
later in this section we compute
$\delta_{\delta g} I_2^{disc}$ rigorously and show that it is a
local anomaly.

\subsubsec{Formal Metric Independence}

Let $\Ivd$ denote $\delta_{\delta g} I_V^{disc}(M,\Ao,g)$.
Using integration by parts formally, we find
$$\eqalign{
 \Ivd &= \delta_{\delta g}c_V \int_{M^V}\TR(L_{tot}^I)	\cr
      &= c_VI \int_{M^V}\TR(\dot L_{tot} L_{tot}^{I-1} )	\cr
      &= c_V I \int_{M^V}\TR( (dB_{tot}) L_{tot}^{I-1}  )	\cr
      &= c_VI \int_{M^V}\TR( B_{tot} d (L_{tot}^{I-1}) )\cr
      &= -c_VI(I-1) \int_{M^V}\TR(B_{tot}\delta^\adp_{tot} L_{tot}^{I-2}) \cr
      &=-c_VI(I-1) \int_{M^V}\TR(\delta^\adp_{tot} B_{tot} L_{tot}^{I-2})
,}\etag\fpfone$$
where $c_V= [(3!)^V (2!)^I V! I!]^{-1}$.
Here, we have used
$$\eqalign{
  dL_{tot} &= -\delta_{tot},\hbox{ and}  		\cr
 \dot L_{tot} & =d B_{tot}
,}\etag\aaa$$
which follow from \pl{1}\ and \pl{4}.

In \aaa,
$$ \delta^\adp_{tot}
 =\Sum_{i\ne j} \delta(x_i,x_j) \delta_{ab}~j^a_{(i)} j^b_{(j)}
.\etag\aab$$
Note that the terms with $i=j$ in \aab\ vanish because
$\delta_{ab}$ is symmetric and the $j$'s are Fermionic.

To show that the last expression in \fpfone\ vanishes, it
is perhaps most expeditious to describe the basic cancellations
in terms of diagrams.  The last expression in \fpfone\ is equal
to a sum over labeled trivalent graphs with a $\delta^\cg$
propagator on the first edge, a $B$ propagator on the second edge,
and an $L$ propagator on all the other edges;
$$\Ivd =
  -\Sum_{\bar\G} c_V \int_{M^V}\TR\left(\delta^\cg_s(x_{in(1)},x_{out(1)})
   B_s(x_{in(2)},x_{out(2)})
    \Pi_{e=3}^V L_s(x_{in(e)},x_{out(e)})\right)
.\etag\pfdiag$$
This can be rewritten as a sum, with suitably combinatorial factors,
of Feynman amplitudes for trivalent graphs with two marked edges.
The amplitude is the same as that for an unmarked graph, except one is to
make an insertion of a $\delta^\cg$ (rather than the propagator $L$)
on one marked edge and an insertion of a $B$ on the other marked edge.
Now, if the $\delta^\cg$ edge connects the points $y$ and $z$,
then by integrating out the $\delta$ function, we find an amplitude
which corresponds to a graph with a four valent vertex inserted.
One obtains, in this manner, all diagrams with one four valent vertex and one
$B$ edge (with the remaining vertices trivalent and the remaining edges
unmarked).  In fact, each diagram of this type is obtained in three
different ways.  So, the Feynman rule associated to the four-valent
vertex will be a sum of three terms.
Figure \cprop\ illustrates the situation.
In the equations below,
we choose to write amplitudes with Lie algebra indices explicit,
rather than imbedded within form notation.
So in Figure \cprop\ we have written the Lie algebra indices that will
be used in the equations below as well as the names of
the positions of the vertices.

The shaded region in each of the diagrams depicted in Figure \cprop\ is the
same except for the location of the external legs.
Let $ W^{cdef}(x_1,x_2,x_3,x_4)$ be the amplitude
for this region when the external legs are at arbitrary
positions $x_1$, ..., $x_4$ as in Figure \dblob.
Then the amplitude for the top left, top middle, and top right diagrams
in Figure \cprop\ are, respectively,
\eqna\thrp
$$\eqalignno{
&\int_{M_y\times M_z} f_{acd} f_{bef} \delta_{ab} \delta(y,z)
W^{cdef}(y,y,z,z),
   & \thrp{.1} \cr
&\int_{M_y\times M_z} f_{acf} f_{bde} \delta_{ab} \delta(y,z)
W^{cdef}(y,z,z,y),
\hbox{ and}
   & \thrp{.2} \cr
&\int_{M_y\times M_z} f_{ace} f_{bdf} \delta_{ab} \delta(y,z) W^{cdef}(y,z,y,z)
   & \thrp{.3}
.}$$
By a more careful analysis, one can check that the
overall combinatorial factors and signs are the same for each of these
diagrams.
Hence the amplitude for the four vertex diagram (IV) is
$\int_{M_x} G_{cdef} W^{cdef}(x,x,x,x)
$,
where the effective Feynman rule at the four valent vertex is
$G_{cdef} =f_{acd} f_{aef} + f_{acf} f_{ade} +f_{ace} f_{adf}
$.
But this vanishes by the Jacobi identity!


For the case $V=2$, the argument above simplifies.
We now provide the details, including the precise combinatorial factors
and signs.
There are only two, 2-loop diagrams, which are both connected.
We call them the dumbbell diagram and the sunset diagram and they
are illustrated in Figure \dumbset.

The Feynman amplitudes, with the correct signs and symmetry factors, are:
$$ I_{dumbbell} =
    -{1\over 8}\int_{M_{x^1}\times M_{x^2}} f_{a^1b^1c^1}f_{a^2b^2c^2}
     L_{a^1c^1}(x_1,x_1) L_{a^2c^2}(x_2,x_2) L_{b^1b^2}(x_1,x_2)
\etag\idumb$$
for the dumbbell diagram; and
$$ I_{sunset} =
    +{1\over 12}\int_{M_{x^1}\times M_{x^2}} f_{a^1b^1c^1}f_{a^2b^2c^2}
     L_{a^1a^2}(x_1,x_2) L_{c^1c^2}(x_1,x_2) L_{b^1b^2}(x_1,x_2)
\etag\isunset$$
for the sunset diagram.
These are the two nonvanishing terms in
$c_2 \int_{M^2}\TR(L_{tot}^3)$.
Repeating the derivation of \fpfone\ at the diagramatic level,
we find
$$ \eqalign{\dot I_{dumbbell}
   = +{1\over 8}\int_{M_{x^1}\times M_{x^2}} &f_{a^1b^1c^1}f_{a^2b^2c^2}
  B_{a^1c^1}(x_1,x_1) L_{a^2c^2}(x_2,x_2) \delta_{b^1b^2}\delta(x_1,x_2)\cr
   &+\hbox{ permutations}
,}\etag\vid$$
and
$$\eqalign{ \dot I_{sunset}
    = -{1\over 12}\int_{M_{x^1}\times M_{x^2}} &f_{a^1b^1c^1}f_{a^2b^2c^2}
     B_{a^1a^2}(x_1,x_2) L_{c^1c^2}(x_1,x_2) \delta_{b^1b^2}\delta(x_1,x_2)\cr
   &+\hbox{ permutations}
.}\etag\vis$$
The permutation terms mean that we should sum over permutations of
$B$, $L$, and $\delta^\cg$.  The terms where the $\delta$ function is
placed on the handle of the dumbbell vanish because $\delta_{ab}
f_{abc}$ vanishes.  Collecting equal terms, relabeling indices, and
integrating out the delta functions, we find
$$ \dot I_{dumbbell}
   =+{1\over 4}\int_{M_{x}} f_{acd} f_{aef} B_{cd}(x,x) L_{ef}(x,x)
,\etag\vidt$$
and
$$\eqalign{ \dot I_{sunset}
  &=+\half\int_{M_{x}} f_{ace} f_{afd} B_{cd}(x,x) L_{ef}(x,x)  \cr
  &=+{1\over 4}\int_{M_x}[f_{ace}f_{afd}-f_{acf}f_{aed}] B_{cd}(x,x)
L_{ef}(x,x)
.}\etag\vis$$
For the last equality, we used antisymmetry of $L_{ef}(x,x)$ under
exchange of $e$ and $f$.
By the Jacobi identity,
$ f_{acd} f_{aef} + f_{ace} f_{afd} + f_{acf} f_{ade} = 0$,
we conclude that
$ \dot I_2 = \dot I_{dumbbell} + \dot I_{sunset}$
vanishes.

\subsubsec{2-loop Anomalies}

To evaluate $\dot I_V$ rigorously, we will replace the formal use of
integration by parts in \fpfone\ by a proper use of Stoke's theorem.

We make the following definitions:
$$\eqalign{
\Delta_{ij} &=\{(x_1,...,x_V)\in M^V; x_i=x_j\}\hbox{ for $i\ne j$, }\cr
\Delta_{tot}&= \union_{i,j=1 \atop i\ne j}^V \Delta_{ij} \hbox{, and}\cr
\cN_\epsilon &=\{(x_1,...,x_V)\in M^V; d(x_i,x_j) < \epsilon
   \hbox{ for some $i$ and $j$}\}
.}\etag\last$$
Note that when $V=2$, $\Delta_{12}$ is $\Delta_{tot}$ and equals the
diagonal in $M_{x_1}\times M_{x_2}$
Moreover, $\cN_\epsilon$ is isomorphic (by the exponential map) to
$N_\epsilon$, the ball bundle of
radius $\epsilon$ in $TM$.  Its boundary, $\del\bar \cN_\epsilon$ in
$M\times M$ is isomorphic to $S_\epsilon$, the sphere bundle of
radius $\epsilon$ in $TM$.

The finiteness result of \S 4 implies
$$ {1\over c_V} I_V^{disc} = \lez
  \int_{M^V-\cN_\epsilon}\Tr(L_{tot}^I)
.\etag\last$$
Hence\foot{
  By generalizing the proof of finiteness in \S 4,
  one can prove that the integral
  $\int_{M^V}\TR(\dot L_{tot} L_{tot}^{I-1})$ converges absolutely.
  This implies that the limits in the last two lines of \fone\
  converge and justifies our exchanging the order of integration
  and metric variation for the second equality of \fone.
},
$$\eqalign{
 {1\over c_V}\Ivd &= \lez
        \delta_{\delta g} \int_{M^V-\cN_\epsilon}\TR(L_{tot}^I)	\cr
      &= \lez I \int_{M^V-\cN_\epsilon}
         \TR(\dot L_{tot} L_{tot}^{I-1} )	\cr
      &= \lez I \int_{M^V-\cN_\epsilon}
         \TR( (dB_{tot}) L_{tot}^{I-1}  )
.}\etag\fone$$
Now using Stoke's theorem on the manifold $M^V-\cN_\epsilon$,
we find
$${1\over c_V}\Ivd = \lez \left[
       I \int_{M^V-\cN_\epsilon}\TR( B_{tot} d (L_{tot}^{I-1}) )
       +I \int_{\del(M^V-\cN_\epsilon)}\TR(B_{tot} L_{tot}^{I-1} ) \right]
.\etag\ftwo$$
But $dL_{tot}$ vanishes away from $\Delta_{tot}$, so the first term
on the right hand side of \ftwo\ vanishes.
Thus we find,
$${1\over c_V}\Ivd = \lez
       I \int_{\del(M^V-\cN_\epsilon)}\TR(B_{tot} L_{tot}^{I-1} )
.\etag\fthree$$

When $V=2$, we obtain
$$\dot I_2 = 3c_2\,\lez
       \int_{(z,u)\in S_\epsilon} \TR( B_{tot}(x,y)  L_{tot}^2(x,y))
.\etag\fit$$
We remind the reader that the $(z,u)$ coordinates are related to
the $x$, $y$ coordinates by the exponential map \expmap.

For the next lemma, it is convenient to define
bounded pieces of propagators:
\eqnn\blbded\eqnn\bbd\
$$\eqalignno{
 L^{bded} &= L^\bd + L^{cont},\hbox{ and} &\blbded\cr
 L_{tot}^{bded}(x,y) &= L_s(x,x) + L_s(y,y) + 2 L_s^{bded}(x,y). &\bbd
}$$
Note also that
$$\eqalign{L_{tot}^{sing}(x,y) & = 2L_s^{sing}(x,y)   \cr
  L_{tot}^\bd (x,y) &= 2L_s^\bd(x,y)
.}\etag\bbe$$
We make similar definitions and observations for $B$.

We now prove
\proclaim{Lemma \newthm\thmlmo}.
$$\dot I_2 = 3c_2\,\lez
       \int_{S_\epsilon}\TR( L_{tot}^{sing}B_{tot}^{bded}  L_{tot}^{bded})
.\etag\lma$$
To do so, we substitute $B_{tot} =B_{tot}^{sing} + B_{tot}^{bded}$
and $L_{tot} = L_{tot}^{sing} + L_{tot}^{bded}$ into \fit\ and expand the
result into eight terms.
The term involving only bounded pieces (i.e. only $L_{tot}^{bded}$ and
$B_{tot}^{bded}$) vanishes because the integrand is bounded and the measure
of the region of integration shrinks to zero as $\epsilon\rightarrow 0$.
The terms involving at least two singular pieces vanishes because
$(L_{tot}^{sing})^2$ and $L_{tot}^{sing} B_{tot}^{sing}$ both vanish.
(Since they are forms of degree greater than $2$ which only depend
on the components of $\tdu$ in the directions orthogonal to $u$.)
Finally, the quantity
$\int_{S_\epsilon}\TR( B_{tot}^{sing} L_{tot}^{bded} L_{tot}^{bded}) $
vanishes in the limit as $\epsilon$ goes to $0$ because
the volume of the area of the sphere $S_\epsilon|_x$
at each point $x\in M$ shrinks like $\epsilon^2$,
while $B_{tot}^\sing$ diverges like $1/\epsilon$,
and $L_{tot}^{bded}$ remains bounded.

Next, we have
\proclaim{Lemma \newthm\thmlmt}.
$$\dot I_2 = 3c_2\,\lez
       \int_{S_\epsilon}\TR( L_{tot}^{sing}B_{tot}^{bd} L_{tot}^{bd})
.\etag\lmtwo$$
To prove the Lemma, we use \blbded\ and expand \lma\ into four terms.
The Lemma follows from
\eqna{\tth}
$$\eqalignno{
0 &=\lez
    \int_{S_\epsilon}\TR( L_{tot}^{sing}B_{tot}^{bd}
L_{tot}^{cont})&\tth{.1}\cr
0 &=\lez
    \int_{S_\epsilon}\TR( L_{tot}^{sing}B_{tot}^{cont}
L_{tot}^{bd})&\tth{.2}\cr
0 &=\lez
    \int_{S_\epsilon}\TR( L_{tot}^{sing}B_{tot}^{cont}
L_{tot}^{cont}).&\tth{.3}
}$$
Now the operation
$\lez\circ \int_{S_\epsilon}\TR$
equals the operation
$-\int_{M_z}\TR \circ \lez \circ \int_{u\in S_\epsilon|_z}$.
The minus sign appears because we've taken the non-standard orientation on
$M\times M$.
So the right hand side of \tth{.1}\ equals
$$ -\int_{M_z} \TR\left( 4\, L_{tot}^{cont}(z,z)
      \lez \int_{u\in S_\epsilon|_z}
       L_s^\sing(x,y) B_s^\bd(x,y) \right)
.\etag\ppro$$
Substituting
the explicit expressions for $L^\sing$ and $B^\bd$ given in
\pl{5}\ and \pl{6}\foot{
  We use orthonormal coordinates on $T_zM$ so
  that $g_{ij}=\delta_{ij}$.
}, \ppro\ equals
$$\eqalign{ -\int_{M_z} \TR\bigg(
  \sum_{i,j} 4[j_{(1)}^a j_{(2)}^a]^2  &L_{tot}^{cont}(z,z) \bigg)
\left[-\half\epsilon^{\nu\rho\sigma}
    (\delta\Gamma_{\alpha\rho\sigma}(z) -
       \half \nabla_\alpha\delta g_{\rho\sigma}(z)) dz^\alpha\right]
\cr
  & \lez \int_{u\in S_\epsilon|_z}
\left[ (-\half \epsilon_{\beta\gamma\delta}
       {u^\beta\over ||u||}{ du^\gamma\over ||u||} {du^\delta\over ||u||})
  {u^\nu\over ||u||} \right]
.}\etag\pptwo$$
The second line in \pptwo\ vanishes since it is
the integral of a linear function over the two-sphere
with its standard volume form.
This proves \tth{.1}.
The proof of \tth{.2}\ is similar.

Define $L^{sing,0}$ so that $L^{sing}_{ab}=\delta_{ab} L^{sing,0}$
and make similar definitions for $L^{bd,0}$ and $B^{bd,0}$.
Note that, when restricted to the fibers of $S_\epsilon$ above any point in
$M$,
 $L^{sing,0}$ is minus $\Omega$, where $\Omega$ is the
spherical volume form of unit area with respect to the standard orientation.

To verify \tth{.3}, we have
$$ \eqalign{
\lez &\int_{S_\epsilon}
      \TR\left( L_{tot}^{sing}B_{tot}^{cont} L_{tot}^{cont} \right)\cr
&= -\int_{M_z}\TR\left(L_{tot}^{cont}(z,z) B_{tot}^{cont}(z,z)
       [j_{(1)}^a j_{(2)}^a] \right)
      \lez \left[\int_{S_\epsilon|_z} 2 L_s^{sing,0}\right]
 \cr
&= 2\int_{M_z}\TR\left(L_{tot}^{cont}(z,z) B_{tot}^{cont}(z,z)
       [j_{(1)}^a j_{(2)}^a] \right)
.}\etag\ll$$
Finally, after unraveling all the notation,
the last term in \ll\ vanishes by the formal proof of
metric independence.

\proclaim{Theorem \newthm\thmtla}.
$$\dot I_2 = -\left({h\dim(G)\over 24}\right){1\over 8\pi^2}
    \int_{M_z} \delta\Gamma_{\alpha\rho}^\nu R_{\beta\gamma}{}^\rho{}_\nu
        dz^\alpha dz^\beta dz^\gamma
,\etag\tft$$
where $h$ equals the dual Coxeter number of $G$.

In the basis of $\cg$ we have chosen, $h$ is given by
$2h dim(G)=f_{abc} f_{abc}$.

Substituting \bbe\ int \lmtwo\ and performing
the $\TR$ operation, we find
$$ \dot I_2 = -{1\over 4} f_{abc} f_{abc}\int_{S_\epsilon}
  L^{sing,0}(x,y) B^{bd,0}(x,y) L^{bd,0}(x,y)
.\etag\last$$
Now substitute in the explicit expressions for $L^{sing}$,
$L^\bd$, and $B^\bd$ and perform  the integral over the sphere:
$$\eqalign{
  \dot I_2 & ={h \dim(G)\over 8(2\pi)^2}
   \int_{M_z}\lez \int_{u\in S_\epsilon|_z}
   \Omega (g_{\mu\nu} {u^\mu\over ||u||} \hat R^\nu)
     (g_{\rho\sigma}{u^\rho\over ||u||}) g_{\rho\sigma} \hat\cO^\sigma  \cr
  &= {h \dim(G)\over 24(2\pi)^2} \int_{M_z}\hat R^\mu \hat O^\nu g_{\mu\nu}
.}\etag\last$$
The final expression equals is a sum of two terms:
the first term is the right hand side of \tft;
the second term vanishes by integration by parts and the Bianchi identity.

\smallskip

Let $\cs_{grav}(g,s)$ be the Chern--Simons action for the metric
connection associated to $g$ defined using a framing $s$ of the manifold $M$
and normalized using minus half the trace in the adjoint representation
of $SO(3)$.
(By framing we mean a homotopy class of trivializations
of the tangent bundle of $M$.)
Since
$$ \delta_{\delta g} \cs_{grav}(g,s)
  = -{1\over 4\pi}
   \int_{M_z} \delta\Gamma_{\alpha\rho}^\nu R_{\beta\gamma}{}^\rho{}_\nu
        dz^\alpha dz^\beta dz^\gamma
,\etag\last$$
we obtain
\proclaim{Cor.~\newthm\thmmfi}.
Let $M$ be an oriented compact $3$-manifold without boundary.
Let $s$ be a framing and
$\Ao$ an acyclic flat $G$-connection.  Then the quantity
$$\tilde I_2 = I_2(M,\Ao,g) - {h\dim(G)\over 24}{1\over 2\pi} \cs_{grav}(g,s)
\etag\tlinv$$
is independent of the metric $g$\foot{
   We warn the reader that we are not confident of the minus sign in the
   Corollary because of the number of sign conventions used in its derivation.
 }.

\newsec\Soutlook{Concluding Remarks}

\subsubsec{I. Shift in $k$ and Higher Loop Anomalies.}

Having described a closed form for the $l$-loop contribution,
and having found a manifold invariant at $2$ loops, we discuss
the relation between our perturbative results so far with Witten's exact
solution.  For a manifold $M$ with framing $s$, the exact solution will be
denoted by $\Zke(M,s)$.

Recall Witten's analysis of the semiclassical approximation,
$Z^{sc}$.  Expanding around a flat acyclic connection $\Ao$, he found that
$$ Z^{sc}(M,\Ao, g) = \tau(M,[A^{(0)}])^\half
                      e^{{i\pi |G|\over 4} \eta_{grav}(M,g)}
                     e^{i(k+h) \cs(\Ao)}
,\etag\last$$
where $\tau(M,[A^{(0)}])$ is the analytic torsion of the flat
connection $\Ao$ on $M$ [\RS], [\Sc],
and $\eta_{grav}(M,g)$ is the $\eta$ invariant of the curl operator on
$M$\foot{
   We normalize $\eta$ invariants as in [\APS].  The
   $\eta$ invariants in [\WiI] are smaller by a factor of two.
}.

As it stands, $Z^{sc}$ is not independent of the metric $g$.
However, from [\APS], one finds
$$ \delta_{\delta g} \eta_{grav}(M,g)
  = -{1\over 6\pi} \delta_{\delta g} \cs_{grav}(g, s)
,\etag\last$$
for any framing $s$
Thus, by adding a local counterterm $|G|/24 \cs_{grav}(g,s)$ to
the classical action, one obtains an invariant
$$ \tilde Z^{sc}(M,\Ao,s) = Z^{sc} e^{{i|G|\over 24} \cs_{grav}(g,s)}
\etag\last$$
of framed manifolds with flat acyclic connection.
Note that the counterterm added is, up to a constant, the counterterm
added in \tlinv\ to get a $2$-loop invariant.

One expects that the leading asymptotics for large $k$ of
Witten's exact solution will reproduce the semi-classical
approximation once the counterterms above have been included.
More specifically, one expects that, asymptotically for large
$k$,
$$ Z_k^{exact}(M,s) \sim \sum_{\Ao} \tilde Z_k^{sc}(M,\Ao,s)
.\etag\cheat$$
This has been confirmed for many examples [\FG], [\Je], [\Gar]\foot{
  In these examples, some of the flat connections have non-zero
  betti numbers $\beta_i$.  In that case, there is
  an extra factor $(k+h)^{\half(\beta_1-\beta_0)}$ in the
  definition of $Z_k^{sc}$.  In order to reproduce the large
  $k$ asymptotics of the exact solution in these examples,
  it is also necessary to include a constant factor of one over the
  order of the center of $G$ whose origin is explained in [\WiIII].
}.
Also, as observed in [\WiI], the behaviour of $Z_k^{sc}$ and
$\Zke$ under a change of framing are consistent with \cheat:
\eqna\framdep
$$\eqalignno{
  \Zke(M,s+1) &= \Zke(M,s) e^{{2\pi i\over 24}{k\over k+h}|G|}&\framdep{.1} \cr
  \tilde Z_k^{sc}    &= \tilde Z_k^{sc} e^{{2\pi i\over 24} |G| }
,&\framdep{.2}
}$$
where we used $\cs_{grav}(g,s+1)=\cs_{grav}(g,s) +2\pi$.

\cheat\ describes the leading asymptotics of $\Zke$.
Ignoring for the moment the integration over the moduli space
of flat connections, one expects for any nice gauge fixing and
regularization procedures
that the asymptotics with higher order terms will be given by
$$ \Zke \sim \tilde Z_k^{perp}
   \mathrel{\mathop=^{\rm def}}\sum_{\Ao} \tilde Z_k^{sc}
      exp\left({\Sum_{l=2}^\infty (-{ik'\over 2\pi})^{1-l}
      \tilde I_l^{conn}}\right)
.\etag\zkas$$
Here $\tilde I_l$ is the regularized contribution of $l$
loop graphs with a counterterm added to insure metric independence,
and $k'$ is a function of $k$ whose form depends on the regularization
scheme.  (There has been much discussion in the physics literature
on this point, see for example [\ALR].)
The only regularization implicit in this paper was (i) summing
over all particle types (BRS ghosts as well as the original dynamical
field) before integrating over $M$, and (ii) point splitting regularization
of the propagator on the diagonal (i.e. the definition of $L_s(x,x)$).

We now argue that $k'=k+h$ for our regularization scheme.
First, we observe that
$$ I_l^{conn}(-M,\Ao,g) = (-1)^{1-l} I_l^{conn}(M,\Ao,g)
\etag\sa$$
where $-M$ is the manifold $M$ with the opposite orientation.
Next we note that the $2$-loop counterterm above changes sign under
orientation reversal.  We expect counterterms at higher loops to
be local and respect the symmetry above, so that
$$ \tilde I_l^{conn}(-M,\Ao) = (-1)^{1-l} \tilde I_l^{conn}(M,\Ao)
.\etag\sb$$

Now let's focus our attention on the case $M=S^3$ with $s$ the framing
invariant under orientation reversing diffeomorphisms.
The exact solution in that case is
$$ \Zke = ({2\over k+h})^\half sin({\pi\over k+h})
.\etag\last$$
Although the only flat connection on $S^3$, the trivial connection,
is not acyclic, we can still show \sb.
Thus $\tilde I_l(S^3,0)$ vanishes for $l$ even, so that
$\zkas$ implies $k'=k+h$.

Next we discuss the specific form of the counterterms, assuming \zkas\
with $k'=k+h$.
Let us assume that
$$ \tilde I_l = I_l + \beta_l \cs_{grav}(g,s) + a_l(g)
,\etag\last$$
where $a_l(g)$ is a local counterterm independent of any choice of framing
and $\beta_l$ is a constant.
We have already seen that this is the case for $l=1,2$, with $a_1=a_2=0$.
Preliminary investigations based on naive power counting support this
assumption and even suggest that $a_l(g)=0$.

The framing dependence
$$ \tilde Z_k^{perp}(M,s+1)
  = \tilde Z_k^{perp}(M,s) e^{2\pi\Sum (-{i(k+h)\over 2\pi})^{1-l} \beta_l}
\etag\last$$
togeteher with \zkas\ and \framdep{.1} imply that
(i) $\beta_1= i|G|/24$.
(ii) $\beta_2= -{h|G|/24}$, and (iii) $\beta_l=0$ for $l>2$.
Item (i) is the agreement of the framing dependence of the
semiclassical approximation with that obtained from the exact solution.
Item (ii) says that the framing dependence of our $2$-loop invariant
given in Cor.~\thmmfi\ agrees with that expected from
the exact solution.  (The caveat to the footnote in Cor.~\thmfin\ holds
here as well.  Other authors seem to have had as much trouble with
sign conventions as we do.)
Finally,  item (iii) implies that $\beta_l=0$.
So, we are hopeful that, for $l>2$, $I_l$ is metric independent.

\subsubsec{II. Extensions.}

In the paragraphs above, we discussed the $l$-loop counterterm and the
shift from $k$ to $k+h$ which needs to be  understood better\foot{
  A derivation was suggested in a talk by one of us (SA) in May, 1991 at
  a Conference on ``Topological and Geometrical Methods in Field Theory'' in
  Turku, Finland.  But evidence against this was subsequently
  supplied by E.~Witten.
}.
We now list other possible extensions of our results and comment on them.

\noindent i.  One should remove the assumption of acyclicity of $\Ao$.
In fact, our proof of finiteness of $I_l(M,\Ao,g)$ remains valid without
the assumption
of acyclicity.  (First, the propagator is still well defined; see the
comment after \dflh.  Second, $L^{had}$ has the same form even when
the cohomology doesn't vanish.)
However, one must show that the integral of $I_V^{disc}(M,\Ao,g)$
over the moduli space of flat connections is finite\foot{
  The integral is relative to $Z_k^{sc}$, which still
  needs to be defined precisely as a measure on the moduli space
  of flat connections, as opposed to a section of a line bundle.
}.
One knows formally that the variation
of $I_V^{disc}$  is a total divergence, so that its integral
is metric independent.  This formal proof must be
made explicit and rigorous up to local anomalies.  Considering the
complicated structure of the moduli space, it seems at present
a formidable task.  We can at least say that Cor.~\thmmfi\ holds
for $M$ a rational homology sphere and $\Ao$ the trivial connection.

\noindent ii.  Our results should be extended to include knot
invariants.  Actually, since we have already shown finiteness of the
integral of Greens functions against a bounded source,
little needs to be added to show that the terms in the perturbative
expansion of Wilson loop expectation values is finite.
The $l$-loop contribution to Wilson loop expectation values
can be expressed in terms of
the untrunctated Greens functions with $E$ external legs.
The latter can be
interpreted as elements of $\Omega^E(M^E;\cg^{\otimes E})$.
Formally, assuming acyclicity of $\Ao$,
these Greens functions are closed, and their derivatives
with respect to a metric variation are exact.  This will lead to
the formal proof of metric independence.

\noindent iii.  It is natural to try to extend
our results to the case when the manifold $M$ has
a boundary.  Now the exact solution produces a state vector
in a Hilbert space associated to the boundary in a nonperturbative
way.  It is not clear, however, where the perturbative terms belong.

Since the the sewing rule is one of the fundamental properties of
a topological field theory reflecting the intuition of path integrals,
one would hope to capture some analogue of it at the perturbative level.
More modestly,
perhaps one could get some feel for factorization perturbatively
by trying to prove directly the sewing formula for connected sums:
$$ \Zke(M_1 \# M_2) = {\Zke(M_1) \Zke(M_2)\over \Zke(S_3) }
.\etag\css$$
That might require computing the perturbative terms for $\Zke(S_3)$
(or anything) directly.

\noindent iv.  To reproduce the full exact solution, rather than
just its asymptotic behavior to all orders,
it would be necessary to sum the perturbation series.
Explicit examples seem to indicate that
the full partition function is analytic in ${1\over k+h}$ near $0$.
\nonpert\ might be useful for showing this.

\noindent v.  Finally, it would be of great interest, for example in the
application to $3$-dimensional gravity, to generalize the results contained
here to non-compact gauge groups.   Hopefully one could extend the $1$-loop
analysis of this problem given in [\BNW] to a sensible regularization to
all orders in perturbation theory.  The manifold invariants thus obtained
would have, in our opinion, a rich geometric interpretation in their own right.

\noindent{\it Acknowledgements: }
We would like to thank Edward Witten and Dror Bar-Natan for useful discussions.

\listrefs

\bye